# Impacts of Surface Depletion on the Plasmonic Properties of Doped Semiconductor Nanocrystals

Omid Zandi[1†], Ankit Agrawal[1†], Alex B. Shearer[1], Lauren C. Gilbert[1], Clayton J. Dahlman[1], Corey M. Staller[1], Delia J. Milliron[1]*

[1] *McKetta Department of Chemical Engineering, The University of Texas at Austin, Austin, Texas 78712, United States*

[†]Equal contribution

*Corresponding author: milliron@che.utexas.edu

**Abstract**

Degenerately doped semiconductor nanocrystals (NCs) exhibit a localized surface plasmon resonance (LSPR) in the infrared range of the electromagnetic spectrum. Unlike metals, semiconductor NCs offer tunable LSPR characteristics enabled by doping, or via electrochemical or photochemical charging. Tuning plasmonic properties through carrier density modulation suggests potential applications in smart optoelectronics, catalysis, and sensing. Here, we elucidate fundamental aspects of LSPR modulation through dynamic carrier density tuning in Sn-doped $In_2O_3$ ($Sn:In_2O_3$) NCs. Monodisperse $Sn:In_2O_3$ NCs with various doping level and sizes were synthesized and assembled in uniform films. NC films were then charged in an *in situ* electrochemical cell and the LSPR modulation spectra were monitored. Based on spectral shifts and intensity modulation of the LSPR, combined with optical modeling, it was found that often-neglected semiconductor properties, specifically band structure modification due to doping and surface states, strongly affect LSPR modulation. Fermi level pinning by surface defect states creates a surface depletion layer that alters the LSPR properties; it determines the extent of LSPR frequency modulation, diminishes the expected near field enhancement, and strongly reduces sensitivity of the LSPR to the surroundings.

Degenerately doped semiconductor nanocrystals (NCs) display unique localized surface plasmon resonance (LSPR) properties, where, unlike metals, the carrier density is tunable over several orders of magnitude ($10^{18}$–$10^{21}$ cm$^{-3}$)[1–3]. Such flexibility in carrier concentrations allows the LSPR frequency ($\omega_{LSPR}$) to be varied broadly from the visible through the mid-infrared (IR) spectrum. Tunability of $\omega_{LSPR}$ originates from excess electronic charge associated with charged defects such as oxygen vacancies[4,5] and aliovalent dopants[6–9], redox and photochemical charging[10,11], or electrochemical modulation[12–15] in NC films. LSPR of metal oxide NCs has been leveraged for numerous applications such as in smart windows[15–17], optoelectronics[18,19], sensing[20], and catalysis[21].

The LSPR in semiconductor NCs is commonly modeled as Drude response of free carriers, as in metals[22–24]. The underlying assumption behind such treatment is that once the semiconductor is degenerately doped, the effects of the band structure on the LSPR properties can be neglected. Although this is a reasonable assumption for describing the dielectric function and associated optical properties of bulk materials, near-surface electronic structure of semiconductors is known to be substantially modified due to the presence of surface defects, surface trap states, and interaction with surface-bound molecules. It has been demonstrated that surface traps modify optical transitions, such as exciton and interband transitions, impacting optical properties such as photoluminescence and two-photon upconversion[25–27]. Native surface states or an externally applied potential can pin the Fermi level at the surface potential leading to the formation of a depletion or accumulation layer near the surface of semiconductor NCs in which the charge carrier density differs markedly from that far from the surface[28,29]. Surface depletion or accumulation results in a spatial gradient in carrier concentration, leading to a spatially varying dielectric function. Such band diagram modifications and their effect on plasmonic properties have been studied in degenerately doped semiconductor thin films such as Sn:In$_2$O$_3$, but their impact has not yet been analyzed in NCs[30]. In NCs, due to high surface to volume ratio, the influence of surface perturbation on LSPR properties is expected to be even more significant.

Colloidally synthesized NCs such as Sn:In$_2$O$_3$[7], CuTe$_{2-x}$[31], or CuS$_{2-x}$[32] are some of the semiconductor systems that have been studied for LSPR based application such as sensing[33], catalysis[21], surface enhanced infrared absorption spectroscopy (SEIRA)[34], and smart windows[15]. We anticipate that the efficacy of NC materials for applications such as smart windows or remote sensing[35], where the dynamic LSPR modulation is achieved by electrochemically or chemically changing the $E_F$, will depend on the extent of surface depletion as a function of NC size and doping level. Contradictory reports in the literature on dynamic LSPR modulation in metal oxide NCs either offer a qualitative description of LSPR frequency modulation based on Drude model[12,36,37], ignoring any depletion effects, or offer a qualitative, simplified description of how depletion could be responsible for minimal shifts of the LSPR frequency[14,38]. Additionally, for applications such as sensing, much lower sensitivity of $\omega_{LSPR}$ to the dielectric constant of the surrounding medium has been observed compared to expected shifts[39]. Qualitatively, the presence of a depletion layer near the NC surface may be expected to reduce the sensitivity of NC LSPR to the changes in the surrounding dielectric environment, but this effect has typically been ignored while analyzing such systems.

In this work, we investigate the effects of surface depletion on the plasmonic properties of doped semiconductor NCs. First, the surface potential, which governs the band bending (band bending potential, $V_{bb}$, is defined as the difference between the flat band Fermi level, $E_{F,fb}$, and the surface potential, $E_s$) was controlled electrochemically and resultant changes in LSPR were monitored using *in situ* Fourier transform infrared (FTIR) spectroscopy. Second, LSPR sensitivity to changes in the surrounding dielectric medium was measured as a function of NC size and dopant concentration. For this purpose, Sn:In$_2$O$_3$, a well-studied plasmonic semiconductor, was chosen as a model system for the following reasons. First, the chemistry of Sn:In$_2$O$_3$ NC synthesis is well established, which enabled us to systematically vary the doping level and size of the NCs. Second, Sn:In$_2$O$_3$ is a technologically relevant material for applications including opto-electronics[18], smart windows[12,40], and sensing[33,34]. Additionally, metal oxide NCs are well known to possess surface defects states due, at least in part, to the presence of surface hydroxyl groups; these defect states define the Fermi level at the surface[14,30]. Using a combination of *in situ* FTIR spectroelectrochemistry (FTIR SEC), solvent dielectric sensitivity assessment, and computational modeling of electron distribution and LSPR modulation, we establish a systematic relationship between the band bending and the LSPR properties of Sn:In$_2$O$_3$ NCs. We find that by varying the size and doping level of the Sn:In$_2$O$_3$ NCs the extent of modulation in the LSPR intensity and frequency can be fully explained by the formation of a dynamic depletion layer near the NC surfaces. Further, we demonstrate, for first time in a plasmonic semiconductor system, that the sensitivity of $\omega_{LSPR}$ to changes in the surrounding dielectric environment tracks with the depletion width, with NCs having the thickest depletion layer being least sensitive to their surroundings. These finding reconcile contradictions in the previous literature regarding LSPR modulation and establishes a general ground to rationally design efficient plasmonic NCs for near-field enhancement, electrochromic, catalysis, and transport applications.

**Results**

The extent of surface depletion (Figure 1a) in semiconductors is determined by the difference between flat bend Fermi level and surface energy ($V_{bb} = E_{F,fb} - E_S$), dopant density, and shape and size of the NCs. Effective control over NC geometry, size, and doping level is thus crucial to make reliable quantitative assessment of the depletion effect on the LSPR. The first objective thus was to reproducibly synthesize highly uniform Sn:In$_2$O$_3$ NCs with various sizes and doping levels. This goal was achieved by adapting a slow growth method recently developed by the Hutchison group[41]. By varying the injection volume and the ratio of In to Sn precursors, a series of monodisperse NCs with variable diameter (7-15 nm) and doping level (1-10% Sn) were synthesized (Figure 1b, Figure S1-S2, and table 1; See methods for synthetic details). The optical extinction spectra of five representative spherical NCs with average diameter of 13.5 nm demonstrate that LSPR blue shifts with increasing dopant density (1-10%) (Figure 1). Fitting of extinction spectrum with Mie theory based on the Drude dielectric function (table S1) indicates that the carrier concentration in the samples varies in between 3.48x 10$^{20}$ and 11.2x10$^{20}$ cm$^{-3}$, increasing with Sn content. The presence of a high initial number of free carriers in our NCs and NC size greater than 7 nm excludes any significant contribution due to quantum effects, thus supporting the continuum analysis and modeling employed later in this work[55,56].

| Sn% in colloidal flask | Sn% estimated by ICP | Size (nm) | Carrier Concentration $10^{20}$(cm$^{-3}$) |
|---|---|---|---|
| 1 | 1.1 | 7.4±1.3 | 3.47 |
| 1 | 1.0 | 15.4±3.1 | 2.90 |
| 2 | 1.8 | 8.7±1.1 | 5.77 |
| 2 | 2.4 | 14.6±2.6 | 4.74 |
| 3 | 2.7 | 8.2±0.9 | 7.77 |
| 3 | 2.8 | 13.3±1.9 | 7.91 |
| 5 | 4.8 | 7.4±1.1 | 10.30 |
| 5 | 3.6 | 12.8±1.7 | 8.56 |
| 10 | 7.2 | 7.4±1.0 | 11.11 |
| 10 | 8.2 | 11.5±1.8 | 10.80 |

**Table 1**. Nominal Sn doping percentage in the colloidal flask, actual Sn doping percentage incorporated into the Sn:In$_2$O$_3$, mean size and standard deviation obtained by analyzing TEM image and NCs drude carrier concentration obtained by fitting experimental extinction spectra.

To study the role of applied potential on LSPR properties, uniform thin films of NCs were spin coated from NC dispersions on conductive Sn:In$_2$O$_3$-coated glass substrates and assembled into a custom-designed multi-layer sandwich cell for *in situ* FTIR SEC measurements (see the methods and Figure S3-S6). The *in situ* electrochemical cell was designed to allow us to monitor FTIR spectra as a function of applied potential while minimizing absorption due to the electrolyte and contacting electrodes.

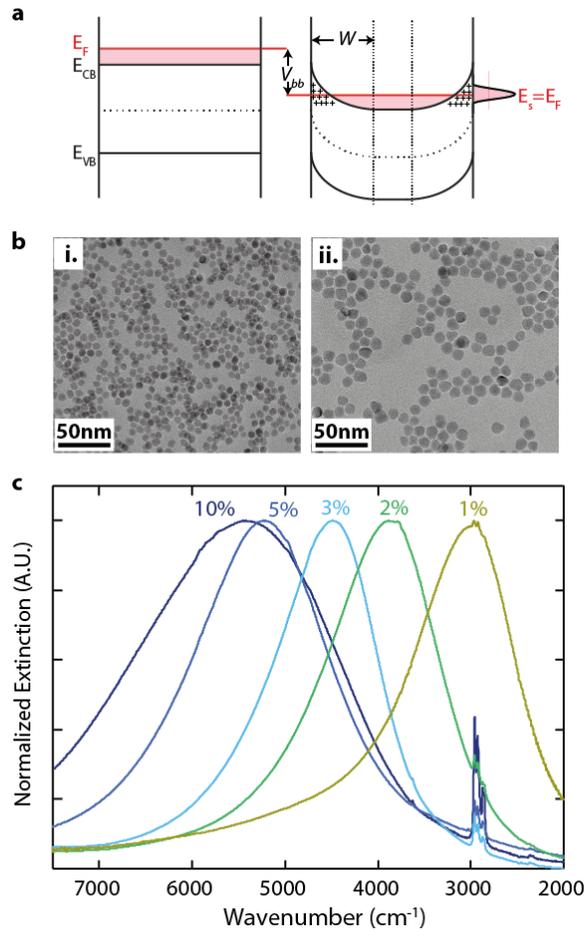

**Figure 1. LSPR in Semiconductor NCs.** a) Schematic representation of the band energetics of a NC under flat band, and band bending conditions where $E_S$ represents the surface state potential. b) Representative TEM images of monodisperse 3% doped $Sn:In_2O_3$ NCs with diameters of i) $8.19 \pm 0.90$ and ii) $13.30 \pm 1.90$ nm. Additional TEM images are provided in the SI. c) LSPR extinction spectra of $Sn:In_2O_3$ NCs with nominal doping levels ranging from 1-10% (dispersed in tetrachloroethylene).

Introducing a larger concentration of Sn dopants in $In_2O_3$ lattice results in significant band gap and band edge modification[42,43]. The interaction among free carriers and with ionized impurities can give rise to a downward shift of the conduction band edge ($E_{CB}$) and an upward shift of the valence band edge ($E_{VB}$), resulting in a net band gap narrowing effect. The optical band gap, on the other hand, increases due to filling of the conduction band states as the $E_F$ is raised (Burstein-Moss shift). The two competing effects determine the dependence of the optical band gap and flat band Fermi level ($E_{F,fb}$) on the Sn doping concentration (Figure 2a)[43]. The final position of the $E_{F,fb}$ determines the extent of the depletion, thus the LSPR modulation ($V_{bb} = E_{F,fb} - E_S$).

The *in situ* LSPR modulation spectra of NCs with two extreme sizes (average diameter 7.4 and 13.4 nm) and doping levels (1% and 10% Sn, see table 1 for the actual doping levels) in response to applied potential are shown in Figure 2b-2e. The LSPR spectra were recorded upon charge saturation after applying a constant potential step for 3 min to insure a steady state carrier concentration throughout the NCs ensemble (Figure S5). Strong modulation of the LSPR extinction intensity, resulting from capacitive charging/discharging of the NCs, was observed in all cases irrespective of NCs size or doping level. This behavior is consistent with previous studies on electrochemical modulation[44] and is analogues to photochemical charging of doped metal oxide NCs[36,37,45]. Unlike the extinction intensity, the $\omega_{LSPR}$ modulation exhibited significant differences for different doping levels and NC sizes. In the case of 1% doped-7.4 nm NCs (Figure 2b.i) increase in the LSPR extinction is accompanied by a significant blue shift of $\omega_{LSPR}$ ($\Delta\omega = 704$ cm$^{-1}$). In contrast, with 10% doped-7.4 nm NCs (Figure 2d.i), the $\omega_{LSPR}$ shift was significantly smaller ($\Delta\omega = 331$ cm$^{-1}$). Moreover, the extent of frequency modulation was not only dependent on the dopant concentration but also on the size of the NCs; $\Delta\omega$ decreased to 424 and 226 cm$^{-1}$ for 15.4 nm NCs doped at 1% and 11.5 nm NCs doped at 10%, respectively (Figure 2c.i, 2e.i). These trends can, at best, only be qualitatively explained with the assumption of uniform increase in carrier concentration throughout the NC[11,37] or formation of an accumulation layer upon application of reducing potential[44]. These observations and inconsistencies in understanding of modulation based of the prior works[14,44] motivated us to explore the effect of surface depletion on LSPR modulation in greater detail. For different NC dopant concentrations and sizes, we numerically solved Poisson's equation as a function of applied surface potential ($E_{app}$) to evaluate radial band profiles. Poisson's equation was solved in spherical coordinates under the parabolic band assumption[46] (see methods, section S3 of the SI and Figure S10-S11). Band energies (Figure 2a) as modified by Sn dopant incorporation into the In$_2$O$_3$ lattice, used in this work for band bending calculations, were referenced from Sn:In$_2$O$_3$ thin film literature[30,43]. For a given surface potential, the depletion width (*W*) is expected to decrease with increasing dopant density and size of the NCs. Since we are solving in spherical geometry, increasing the size of the NC at a given dopant concentration, decreases *W* only slightly, but it substantially decreases the ratio of depleted volume to total nanocrystal volume ($V_w$). $V_w$ dependence on size and dopant density is discussed in detail later in this paper. Solution to Poisson's equation shows that at the most oxidizing $E_{app}$ (0.8 eV) *W*, in the case of 1% doped NCs (Figure 2b.ii and 2c.ii), is around 4 nm, which decreases to less than 1 nm for 10% doped NCs (Figure 2d.ii and 2e.ii). *W* extends through the majority of the NC volume for low doped and small NCs. In contrast, for the high doped case, for all NC sizes (6 -14 nm), it is limited to less than 1 nm.

Applying a reducing potential decreases *W*, with consequences for the spatial distribution of the accumulated charge that depends on both size and doping level. For low doped-6 nm NCs – in which *W* initially extends throughout the NC volume – a reducing potential initially results in charging the entire NC and once the *W* becomes less than the radius of the NC further reduction continues to diminish *W*, thereby increasing the volume of an un-depleted core (Figure 2b.ii). Carrier concentration profiles based on Poission's equation illustrate these two charging regimes for 1% doped-6 nm NC (Figure 4b.i). Substantial changes in carrier concentration occur throughout the NC as $E_{app}$ is varied, explaining the large observed shift in $\omega_{LSPR}$ for NCs like these (Figure 2b.i). For larger NCs (14 nm), where depletion width is smaller than the radius even before adding charge, charging occurs entirely by the second process wherein the volume of the un-depleted core increases (Figure 2c.ii). The carrier concentration remains effectively constant within the un-depleted core while $V_w$ increases substantially, by 87.5%. Therefore, $\omega_{LSPR}$ shift for 14 nm NCs is less than for 6 nm NCs, owing to a smaller change in carrier concentration. The observed blue shift for larger NCs (Figure 2c.i) can be ascribed, in part, to reduction of refractive index in the depleted layer as potential is shifted negatively (Figure S12a). Considering the change in the number of carriers implied by the observed modulation intensity, an equal change in carrier concentration would result in a significantly larger shift in $\omega_{LSPR}$ (Figure S13). Hence, a change in effective volume of plasmonic material must be invoked to explain the modest *Δω* and large intensity modulation.

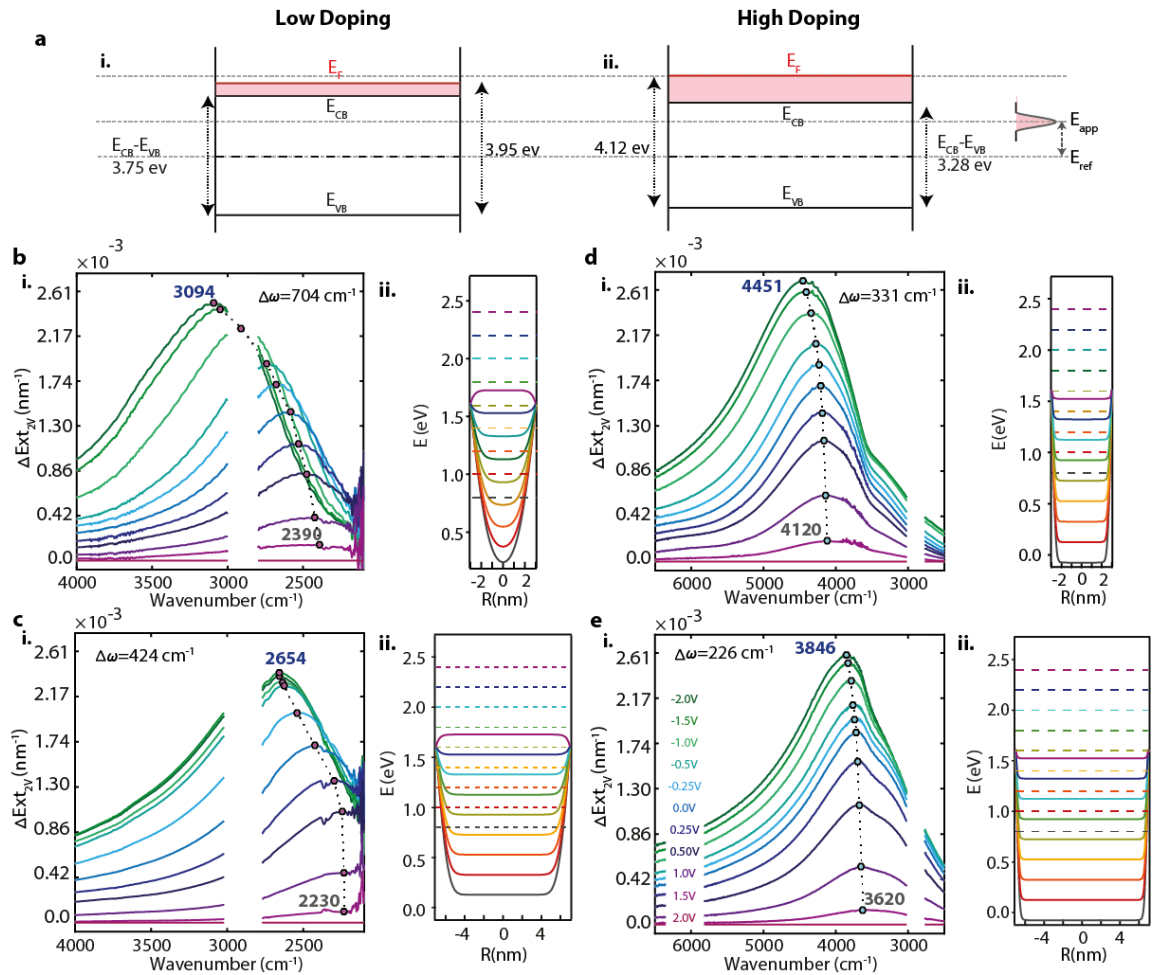

**Figure 2. Electrochemical LSPR Modulation.** a) Band energetics of low and highly doped NCs showing the band gap narrowing and Burstein-Moss effects. b.i-e.i) *In situ* FTIR SEC spectra of a film of NCs collected at various applied potentials for 1%-7.4 nm (b.i), 1%-15.4 nm (c.i), 10%-7.4 nm (d.i), and 10%-11.5 nm (e.i) $Sn:In_2O_3$ NCs. b.ii-e.ii) calculated conduction band profiles corresponding to the case of low doped-6 nm (b.ii) and 14 nm (c.ii) and high doped-6 nm (d.ii) and 14 nm (e.ii) $Sn:In_2O_3$ NCs. Overlapped with the LSPR spectra are the peak positions extracted from fitting the individual spectra. Δω represents the maximum LSPR peak modulation going from +1.5 to -2 V. Electrochemical LSPR modulation was performed by charging the NC films by applying a more reducing potential ($E_{app}$) relative to the reference potential ($E_{ref}$). Note that $E_{ref}$, in the case of experimental results, is the counter electrode potential (*i.e.*, Pt) and in the case of modeling results was taken as the mid point of the band gap where $E_{ref}$=0. Spectral bands where the absorption was saturated due to electrolyte absorption are shown as blank regions.

In highly doped NCs (e.g., 10% Sn, Figure 2d.ii and 2e.ii), $W$ is less than 1 nm at any $E_{app}$, so the charge carriers are consistently delocalized throughout the majority of the NC volume. However, the narrow depletion layer still consumes most of the injected electrons as these NCs are charged, resulting in negligible increase in the electron concentration in the core. The relative change in $V_w$ is 70.4 % and 42 % for 6 nm and 14 nm NCs, respectively, comparing the most oxidized to the most reduced charge state (Figure 2d.ii, 2e.ii). Again, a small shift in $\omega_{LSPR}$ can be ascribed to the dielectric change in the depletion layer upon charging (Figure S12b).

Now we take a closer look at the LSPR modulation dependence on the size and doping level, as increases in either would reduce the relative extent of $V_W$. To probe size effects, three sizes of NCs at each of two extreme doping levels (1 and 10%) were prepared and their *in situ* FTIR SEC spectra were measured. The change in LSPR peak frequency ($\Delta\omega_{LSPR}$) at different potentials versus that at 1.5 V (oxidized state) is shown in Figure 3 (full spectra are provided in Figures S7-S8). The $\omega_{LSPR}$ of low-doped NCs 1% Sn shifts further to the blue upon charging, with the smallest (7.4 nm) NCs exhibiting the highest $\Delta\omega_{LSPR}$ of 704 cm$^{-1}$. As NC size increases, the depletion region occupies a smaller fraction of the NC volume, resulting in smaller $\omega_{LSPR}$ shifts. Likewise, for highly doped NCs (10% Sn) the largest $\Delta\omega_{LSPR}$ is observed for the smaller, 7.4 nm NCs.

To test the effect of the dopant concentration for a given size, a series of NCs with average size of 7.8 nm and variable doping densities ranging from 1 to 10% were prepared. The $V_w$ thus the extent of LSPR frequency modulation, decreases with increasing dopant concentration. The compression of the depletion layer systematically reduces $\Delta\omega_{LSPR}$ as dopant density increases (Figure 3 and Figure S9).

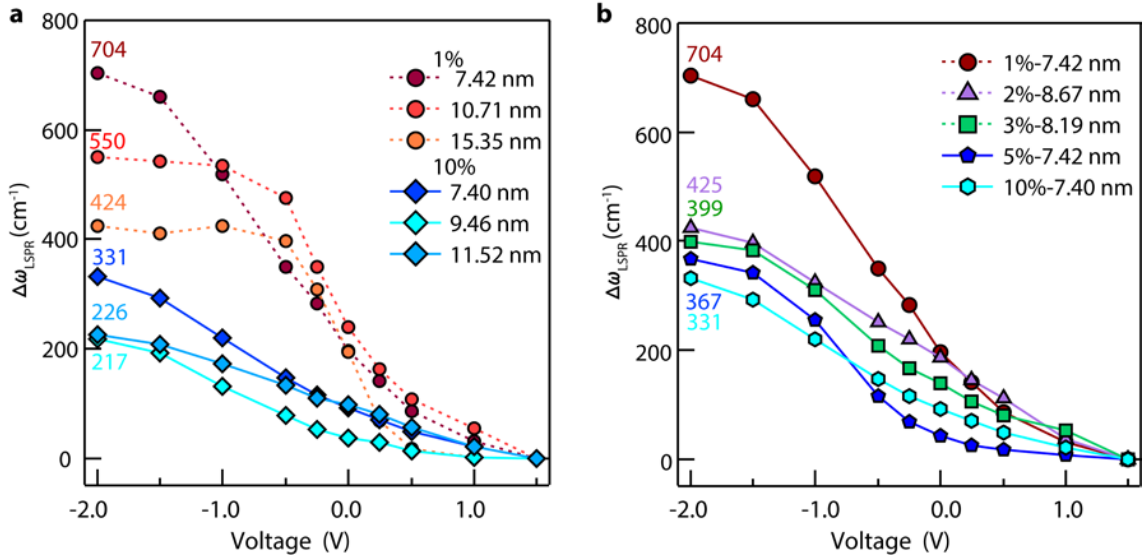

**Figure 3. Effect of NC Size and Doping Concentration on Electrochemical LSPR modulation.** Experimental observations of LSPR peak frequency at different potentials versus that at 1.5 V ($\Delta\omega_{LSPR}$) for Sn:In$_2$O$_3$ NCs with different sizes (a) and doping concentrations (b).

To establish a direct correlation between calculated carrier concentration profiles and optical properties, we developed a multi-scale model to compute optical properties of individual NCs and of films in the *in situ* device. The multi-scale optical model allows us to directly relate the expected optical modulation in a device, similar to our experimental design, to the applied potential. Optical properties of NCs with a given radial distribution of carrier concentration were modeled using an effective medium theory (Figure 4a). The effective dielectric function of a NC ($\epsilon_{eff}$), was calculated using an iterative core-shell effective medium method (See section S5 of the SI). Importantly, we emphasize that effective permittivity calculated using the core-shell model is not equivalent to a Drude type permittivity and cannot be expressed using an effective carrier concentration. (see Section S7 for more discussion on validity of the core-shell optical model). We note that the LSPR behavior slightly changes going from freestanding to assembled NCs due to NC-NC coupling in the film, as well as the convoluted substrate effects. To accurately model the optical property of NCs packed randomly in a thin film, using the $\epsilon_{eff}$ as the dielectric function of an individual NC, the effective film dielectric function, $\epsilon_{eff\_film}$, was calculated (Validity of Maxwell-Garnet approximation is discussed in Section S7). Furthermore, taking into account the substrate and electrolyte effects, the full device stack (based on the experimental system, Figure 4a-ii) transmission and the reflection spectra were then calculated using the T matrix method employing the SCOUT optical package[22] (www.wthesis.com) (See methods, section S6 and Figure S15 and S16 in the SI). Notably, the intensity and frequency modulation of the LSPR predicted by the model (Figure 4 and Figure S19-S21) are in excellent agreement with observed experimental values. The experimental and modeling results compare well across the full range of sizes and doping concentrations. This analysis supports our hypothesis that surface depletion plays a crucial role in determining the LSPR optical properties and their modulation in response to electrochemical charging of NC films.

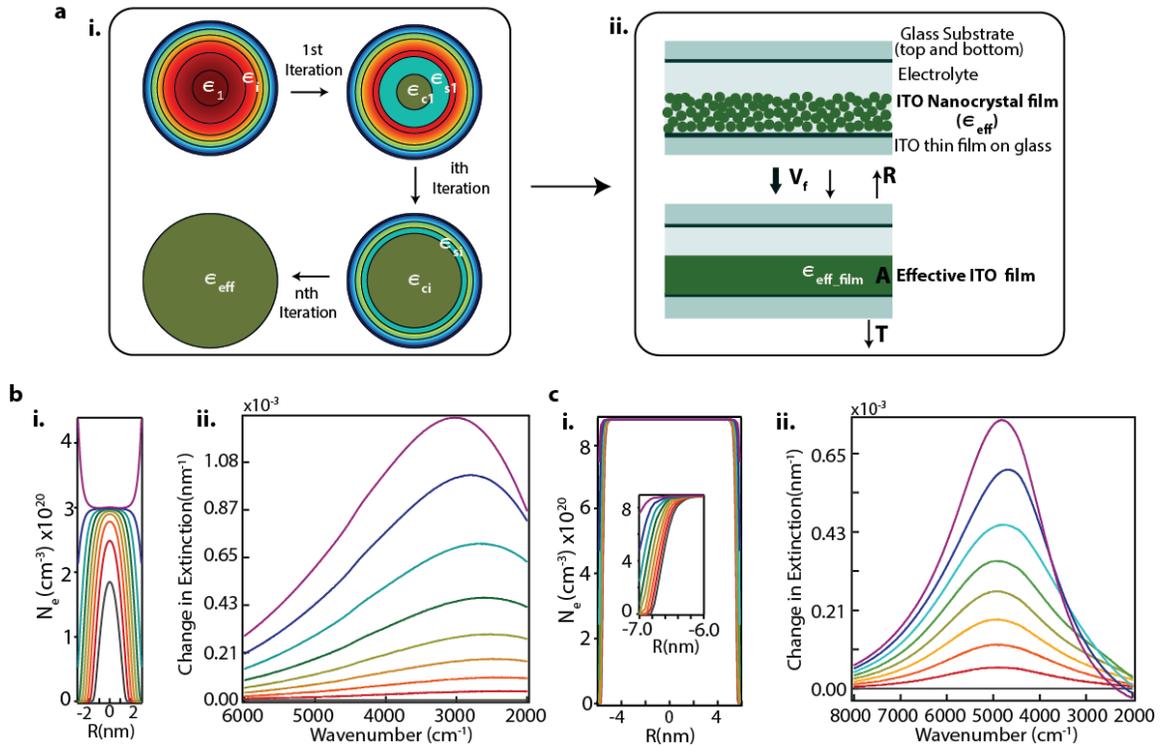

**Figure 4. Optical Modeling of Electrochemical LSPR Modulation.** a) Schematic iterative procedure used to obtain the effective dielectric function of a surface depleted $Sn:In_2O_3$ NC (i) and modeled sandwich cell configuration where $Sn:In_2O_3$ NCs with effective dielectric function, $\varepsilon_{eff}$ are coated on top of an $Sn:In_2O_3$ thin film-coated glass substrate (ii, not to scale). To model the $Sn:In_2O_3$ NC film optical properties, the effective dielectric function of the film was calculated using the Maxwell-Garnett equation. b,c) Calculated carrier concentration profiles (i.) and corresponding LSPR spectra (ii.) for $Sn:In_2O_3$ NC films containing 1% doped-6 nm (b) and 10% doped-14nm (c) NCs.

The correspondence between the shifts of $\omega_{LSPR}$ in experimental and simulated thin films is a strong indication that our depletion layer model captures the relevant physics of LSPR modulation. As a result, we can use simulation to study the LSPR modulation process at the single NC level, isolating the NC from substrate- or NC-NC coupling effects. Computational analysis of optical modulation on the single NC scale (which is not accessible experimentally) will guide future photochemical or electrochemical modulation studies of free standing NCs. For all sizes and doping levels, $\omega_{LSPR}$ shifts of isolated NCs are larger than those in assembled films (Figure 5 and Figure S19-21). Modeling also reveals the accessible range of intensity modulation. For low doping, the LSPR intensity of the depleted (oxidized) NCs is very low, the intensity increases continuously with reducing potential (Figure S19). In contrast, for highly doped NCs, even the oxidized NCs retain substantial LSPR due to the presence of a significant undepleted core (Figure S20). Simulated single NC optical spectra clearly show that the dynamic range for modulating LSPR intensity and frequency is maximized for small sized NCs with a low doping level due to the large spatial extent of the depletion layer compared to the NC size; the modulation range is reduced with increasing dopant concentration and size (Figure S19-S21).

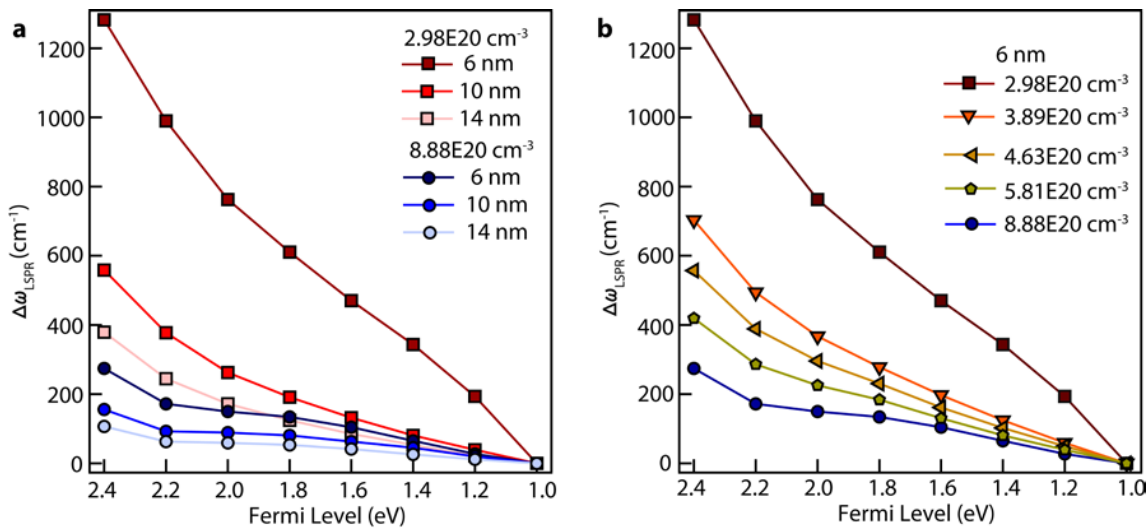

**Figure 5. Modeled size and doping effects on the $\Delta\omega_{LSPR}$ of an isolated NC.** LSPR peak frequency at different potentials versus that at the most oxidized potential (0.8 V) for $Sn:In_2O_3$ NCs with different sizes (a) and doping concentrations (b). Modeled shift in isolated NC, de-convoluted to remove the in-film coupling effects shows greater modulation compared to the corresponding film modulation. Modeled film data, full LSPR spectra and the details of the modeling are provided in the SI.

In the data presented above the depletion layer was manipulated through application of an externally applied potential. In free standing NCs, and in the absence of a well-defined external potential, the Fermi level at the surface is determined by the surface redox chemistry, *i.e.*, the energetics of the surface states. Surface states energetically located below the Fermi level induce a depletion layer the extent of which would depend on the energy difference between $E_F$ and the surface states' energy (Figure 2a). A depleted surface region can be envisioned as a shell of a high dielectric constant material, creating a separation between the active plasmonic core and the surrounding medium. This spatial separation is expected to reduce the sensitivity of $\omega_{LSPR}$ to changes in the medium dielectric constant. Related phenomena have been reported in ligand capped or core-shell metal-dielectric systems[20,47,48]. The sensitivity (*S*) of NC LSPR wavelength ($\lambda_{LSPR}$) to the changes in the surrounding refractive index unit (RIU) is defined as $(\lambda_{LSPR}/RIU)^2$ (see section S9 of the SI for details)[49]. To evaluate NC plasmon sensitivity, NCs with various sizes and doping concentrations were dispersed in solvents with refractive indices raging from 1.3-1.55. From the relationship $\lambda_{LSPR}^2 = Sn^2$ the sensitivity values were extracted and are plotted in Figure 6a for two different sizes and doping levels from 1-10%. As discussed above, the depletion width is higher for smaller NC and lower doping levels. Therefore, larger sensitivity deviation from the case of uniform carrier distribution is expected for the smaller lightly doped NCs. Moreover, for low dopant concentration, there is maximum difference in surface depletion profile for different sizes; as expected this is mirrored in the sensitivity factor (Figure 6a) where the highest deviation from the ideal case is seen for 1%-7 nm. As the carrier concentration increases, dependence of sensitivity factor on NC size reduces, reaching its minimum at the highest doping level (10% Sn) where the *W* is minimized. Thus, the sensitivity perfectly tracks the trend of depletion width, and corroborates the LSPR modulation results discussed above. These results were further confirmed by calculating plasmon sensitivity for NCs of different size and carrier concentrations considering the depletion effect. The results shown in

Figure 6b show systematic size and doping effects on the plasmon sensitivity, in excellent agreement with the experimental observations.

Enhancement of the electric field in nanoscale space surrounding metallic nanoparticles is an important effect of LSPR excitation that enables many emerging applications[50]. We hypothesized that surface depletion would significantly influence the near-field enhancement around plasmonic $Sn:In_2O_3$ NCs. To evaluate such effects, we calculated the near-field enhancement for 1% doped- 6 nm and 12 nm NCs (Section S10 in the SI). The low 1% doping level was chosen as a test case as it allows the strongest LSPR modulation, exhibits strong size dependence, and leads to the greatest deviations from ideal sensitivity. Using the radial profile of the dielectric function calculated based on carrier concentrations obtained from Poisson's equation, we numerically solved for electric field using a finite element method (See section S8 of the SI for more information). Near field enhancement was maximum and size-invariant for the ideal case of un-depleted NCs (Figure 6c). The presence of a depletion layer reduces the near field enhancement surrounding the NC, which underlines our earlier observation of reduced sensitivity to the surroundings for such NCs. The near-field intensity enhancement surrounding a 6 nm NC with a 2 nm depletion layer was reduced by 40% compared to the ideal case, while a 14 nm NC with a 2 nm depletion layer had 26 % lower near-field enhancement than if depletion were absent. These results underscore the strong influence that depletion has on near-field enhancement, especially as the depletion width becomes comparable to NC size, depleting even the core. Furthermore, depletion introduces a size-dependence to the magnitude of near-field enhancement that is otherwise absent (within the quasi-static limit). The additional presence of near field localization within the depleted layer may have other implications for optical properties. All of these deviations from ideal expectations of near-field enhancement are expected to influence the interaction of plasmonic NCs with their environment, including coupling to other NCs or oscillators in the vicinity[50]. This aspect of the current study needs further investigation and is a subject of our current on-going work.

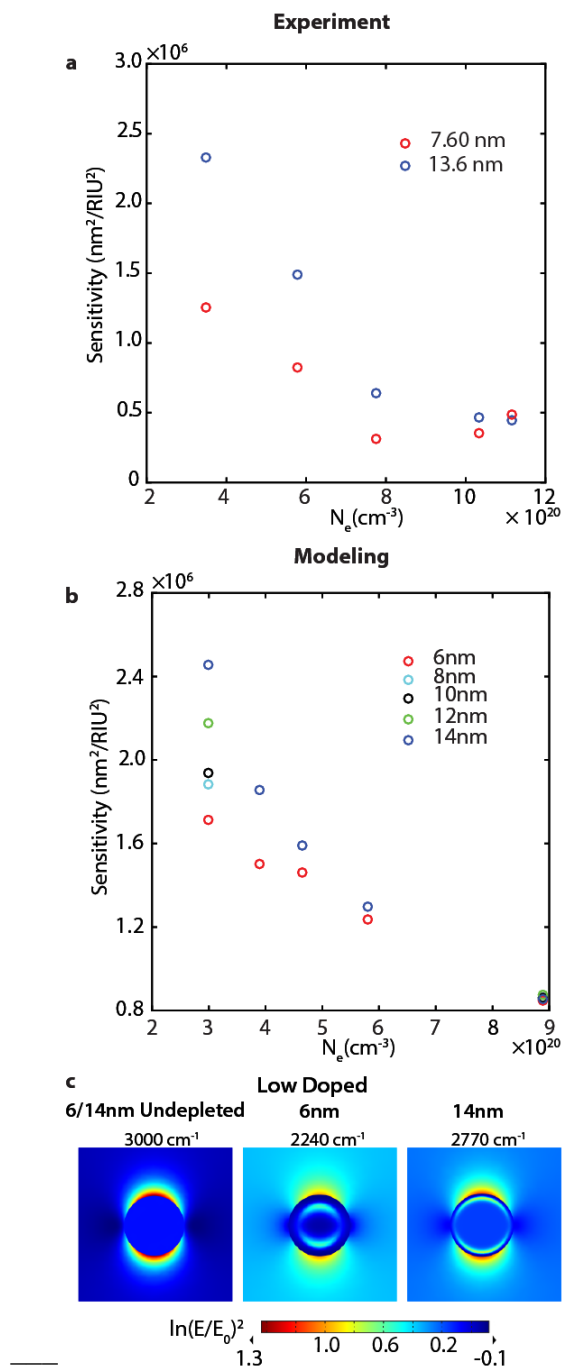

**Figure 6. NC plasmon Sensitivity.** a) Plasmon sensitivity of small and large $Sn:In_2O_3$ NCs with nominal doping concentrations ranging from 1-10%. The sensitivity factor is calculated based on the LSPR peak position in solvent with refractive indices ranging form 1.30-1.55. b) Calculated plasmon sensitivity for $Sn:In_2O_3$ NCs with various sizes and dopant concentrations. c) Calculated near field enhancement maps of a spherical 1% doped $Sn:In_2O_3$ NC showing the effect of surface depletion. Near field enhancement is maximum for undepleted ideal NCs, which decreases for 6 nm and 14 nm surface-

depleted NCs. Near field enhancement decreases substantially for 6 nm NC as the depletion extends through the core of the NC.

These corroborating experimental and modeling results highlight the impact of surface depletion on the plasmonic properties of doped semiconductor NCs. Here, the Fermi level was adjusted by applying a constant potential to a film of $Sn:In_2O_3$ NCs. This allowed us to systematically vary the band bending ($V_{bb}$) and monitor its effect on electron distribution and thus the LSPR behavior. In the absence of the applied external potential the Fermi level is defined by the surface redox potential (due to intrinsic surface states or a deliberately added redox couple[28,37,51,52]). Regardless of their nature, surface states energetically located below the Fermi level induce a depletion layer at the NC surface (Figure 2a). Surface depletion and associated band bending in NCs strongly impacts their potential applications in sensing, electro-optics, and transparent conductors. Surface depletion could also strongly impact carrier transport through NC-based transparent conductors by imposing a potential barrier at NC-NC interface[53].

Strong dependence of LSPR modulation, sensitivity to the surrounding, and near-field enhancement on size and doping concentration of the NCs – all due to surface depletion – highlights the importance of rational choice of the NCs for a given application. For smart window applications[12,15], where the maximum LSPR frequency and intensity modulation is desirable, low doped and small sized NCs are preferable. In contrast, for application such as photon upconversion[54], SEIRA[34] and sensing[48], where the efficacy depends on LSPR coupling to other optical elements in the vicinity of the NC, surface depletion needs to be minimized. Furthermore, the ability to modulate the sensitivity, near field enhancement, and LSPR frequency as a function of surface potential, holds great promise for a range of applications such as electro-optical modulators as well as tunable SEIRA and sensing substrates.

## Methods

**Nanocrystal synthesis.** Sn-doped $In_2O_3$ nanocrystals (NCs) were synthesized using a slow injection method (slow growth) recently reported by the Hutchison group[41]. Briefly, a desired composition of $In(ac)_3$ and $Sn(ac)_4$ (0.5 M total) were dissolved in 7.5 mL of oleic acid at 150 °C for 2 h. In another flask 13 mL of oleyl alcohol was heated to 290 °C under a gentle stream of $N_2$. A desired volume of metal oleate solution was slowly injected to the oleyl alcohol bath using an automated syringe pump at the rate of 0.2 mL/min. Upon injection, the alcohol bath turned light green-bluish in color with the color intensity dependent on the dopant level. After injection, the reaction was held at 290 °C for 20 min before quenching to room temperature. The NCs were washed multiple times using hexane/ethanol solvent/anti-solvent mixtures and finally dispersed in hexane for further use.

The NCs nominal doping level was adjusted by varying the Sn/In ratio in the precursor solution while the NC size was controlled by varying the injection volume. Typically, a 1 mL injection volume resulted in NCs of about 7-8 nm in diameter while for larger NCs (~13-14 nm) a 4 mL injection was required. The size of the NCs were analyzed from TEM images by analyzing at least 200 particles (Figure S1).

**X-ray diffraction (XRD).** XRD patterns were collected for Sn-doped $In_2O_3$ NCs using a Rigako Miniflex 600 with a Cu K alpha radiation source. Two representative patterns for 1 and 10% doped NCs are shown in Figure S2. All the diffraction peaks can be indexed to the reference pattern of the bixbyite $In_2O_3$ crystal structure (PDF# 06-0416).

**Transmission Electron Microscopy (TEM).** TEM images of Sn-doped $In_2O_3$ NCs were acquired using a JEOL-2010F microscope with a Schottky field emission gun and a CCD camera operated at 200kV.

**Inductively Coupled Plasma Mass Spectrometry (ICP-MS).** NC dopant concentration was measured using ICP-AES on a Varian 720-ES ICP Optical Emission Spectrometer. Samples were prepared by allowing NC solutions to dry before digestion in 70% (v) nitric acid. Samples and standards were run in 2% (v) nitric acid.

**Nanocrystal Film preparation**. Uniform thin films of Sn-doped $In_2O_3$ NCs were deposited on commercial glass with a $Sn:In_2O_3$ coating (50 nm) for *in situ* spectroelectrochemical measurements, as well as on Si chips for characterization, using a spin-coating method. For spin-coating, substrates were covered with an adequate volume of NCs dispersion in hexane/octane (1:1 V/V) and spun at 2000 rpm for 1 min, followed by 3000 rpm for 30 s. The NCs assemblies were further ligand exchanged with formic acid driven by mass action; films were immersed in 0.1 M formic acid in MeCN for 30 min followed by rinsing multiple times with a mixture of 1:1 MeCN/CHCl$_3$. At the next step, the formic acid ligands were removed by annealing the films at 120 °C for 20 min in a tube furnace under 50 sccm flow of Ar.

**Device assembly.** The NCs films were assembled in a novel *in situ* FTIR spectroelectrochemistry (SEC) device for SEC measurements. The devices were composed of a working electrode (the Sn-doped In$_2$O$_3$ NC film) and a counter electrode consisting of a glass/ITO substrate coated with Pt nanoclusters. Pt clusters, which served to increase the capacitance of the counter electrode, were deposited by drop-casting Platisol (Solaronix) on the substrates followed by heating in air at 350 °C for 20 min. Electrodes were fused together using a thermoforming Surlyn film (60 μm, Solaronix). A 1.0 cm$^2$ hole in the Surlyn spacer provided a compartment which was further filled with electrolyte (1.0 M tetramethylammonium bis(trifluoromethanesulfonyl)imide (TBATFSI) in tetraglyme). A photograph of the final device is shown in Figure S3.

**Electrochemistry.** Electrochemical measurements were made using a S200 Biologic potentiostat. Three cycles of cyclic voltammetry (CV) were run before the *in situ* FTIR measurements. Capacitive CV profiles indicated the high capacitance of the working and counter electrodes (Figure S4a).

**FTIR Spectroscopy.** Optical transmission spectra were measured using a Bruker Vertex 70 FTIR. For *in situ* FTIR SEC, the device was attached to a holder with an aperture of about 1.1 cm$^2$. The device was titled to a 20° angle to avoid the internal refection patterns from reaching the detector. A typical *in situ* FTIR spectrum was acquired by backgrounding the detector versus ambient air, then applying a constant potential for 3 min before recording the spectra.

**Modeling- Poisson's Equation.** Poisson's equation was solved numerically for spherical NCs by assuming uniform dopant concentration and surface potential, $E_S$, using a finite element method. The charge density at any point inside the NC is made up of mobile electrons, holes and immobile ionized impurity centers. Poisson's equation (Equation S6) was solved in COMSOL using a finite element scheme.

**Modeling- Effective film dielectric function and device optical modeling.** The sandwich cell designed in-house for *in situ* SEC was modeled using commercial SCOUT software ([www.wtheiss.com](www.wtheiss.com)) by employing several built-in optical models, including the T-matrix method to calculate extinction spectrum and the Drude model to express the dielectric function of ITO thin films on a glass substrate.

## Acknowledgement

This research was supported by the National Science Foundation (NSF, CHE-1609656) and the Welch Foundation (F-1848).

## Author Contributions

O.Z. and A.A. contributed equally to this work. O.Z. synthesized the materials, fabricated devices and performed experimental FTIR spectroscopy, A.A. performed simulations, A.S. synthesized $Sn:In_2O_3$ nanocrystals, L.G. performed analysis of TEM images, C.J.D. provided critical conceptual inputs, C.M.S. performed ICP, D.J.M. provided overall guidance, and O.Z., A.A., and D.J.M. wrote the manuscript with critical input from all the authors.

## Competing Financial Interests

D.J.M. has a financial interest in Heliotrope Technologies, a company pursuing commercialization of electrochromic devices.

## Materials and Correspondence

Contact authors: D.J.M. (milliron@che.utexas.edu).

*Supporting Information for*

# Impact of Surface Depletion on the Plasmonic Properties of Doped Semiconductor Nanocrystals

Omid Zandi[†], Ankit Agrawal[†], Alex B. Shearer, Lauren C. Gilbert, Clayton J. Dahlman, Corey M. Staller, Delia J. Milliron*

### Section S1. Synthetic Protocols and Characterizations

**Synthesis Protocol.** Sn-doped $In_2O_3$ nanocrystals (NCs) were synthesized using a slow injection method (slow growth) recently reported by the Hutchison group[1]. Briefly, a desired composition of $In(ac)_3$ and $Sn(ac)_4$ (0.5 M total) were dissolved in 7.5 mL of oleic acid at 150 °C for 2h. In another flask 13 mL of oleyl alcohol was heated to 290 °C under a gentle stream of $N_2$. A desired volume of metal oleate solution was slowly injected to the oleyl alcohol bath using an automated syringe pump at the rate of 0.2 mL/min. Three needles (each packed with a piece Kimwipe) were inserted in one of the rubber stoppers as gas outlet. Upon injection, the alcohol bath turned light green-bluish color with the color intensity dependent on the dopant level. After injection, the reaction was held at 290 °C for 20 min before quenching to room temperature. The NCs were washed multiple times using hexane/ethanol solvent/anti-solvent mixtures and finally dispersed in hexane for further use.

The NCs nominal doping level was adjusted by varying the Sn/In ratio in the precursor solution while the NC size was controlled by varying the injection volume. Typically, a 1 mL injection volume resulted in NCs of about 7-8 nm in diameter while for larger NCs (~13-14 nm) a 4 mL injection was required. The size of the NCs were analyzed from TEM images by analyzing at least 200 particles (Figure S1).

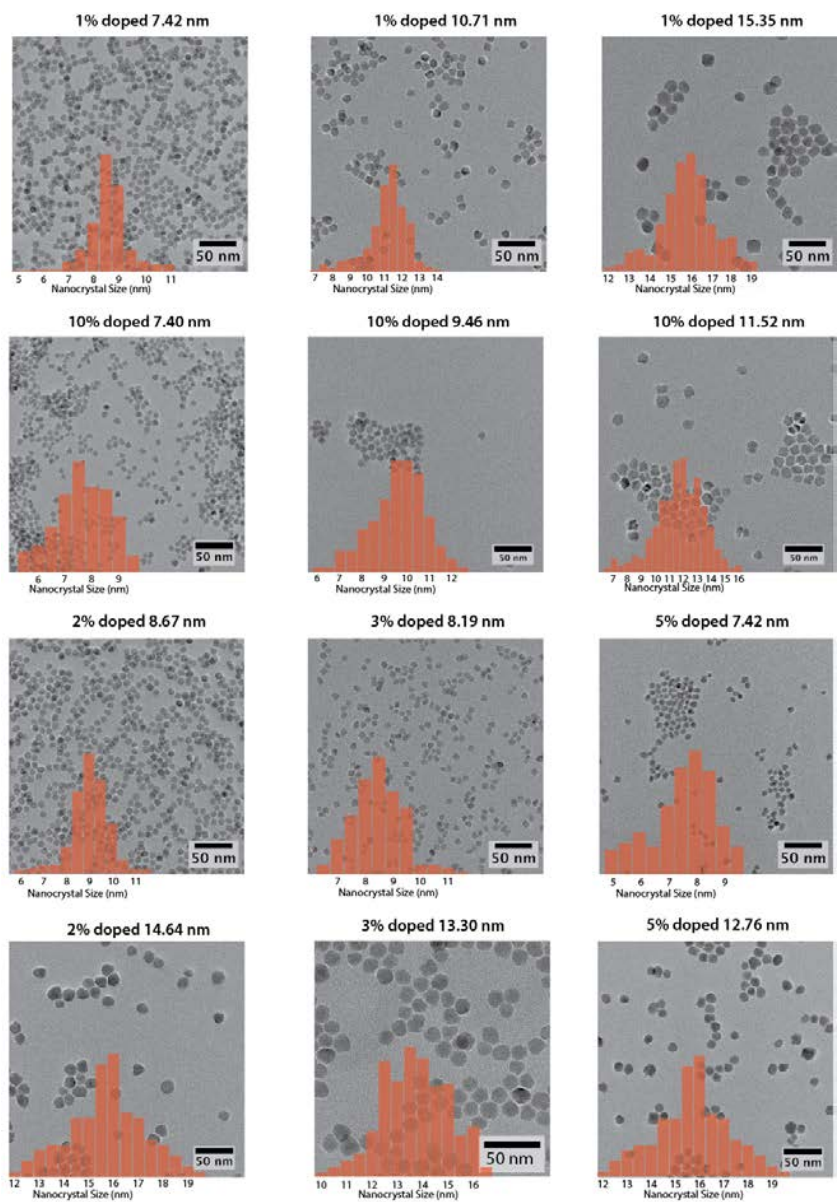

**Figure S1.** TEM images of NCs were acquired using a JEOL-2010F microscope with a Schottky field emission gun and a CCD camera operated at 200kV. For each sample, one representative micrograph is shown, with an overlaid histogram of the sizes for all the analyzed NCs.

**X-ray diffraction (XRD).** XRD patterns were collected for $Sn:In_2O_3$ NCs using a Rigako Miniflex 600 with a Cu K alpha radiation source. Two representative patterns for 1 and 10% doped NCs are shown in Figure S2. All the diffraction peaks can be indexed to the reference pattern of the bixbyite $In_2O_3$ crystal structure (PDF# 06-0416).

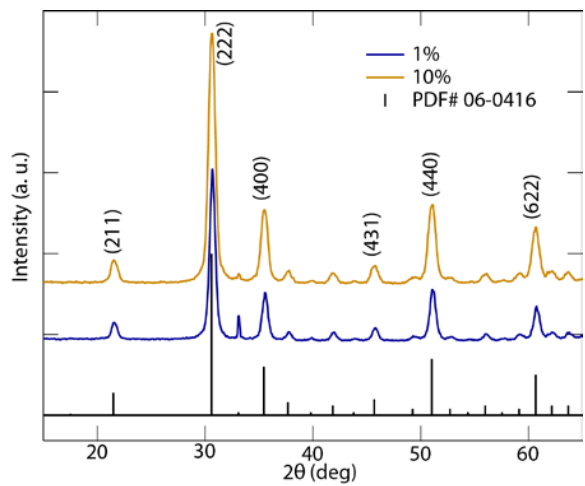

**Figure S2.** XRD diffraction patterns of 1%- 15.4nm and 10%-11.5 nm doped Sn:In$_2$O$_3$ NCs.

**Table S1.** Drude optical parameter and carrier concentration from fitting experimental extinction spectra.

| Doping level % | Size (nm) | Plasma Frequency (cm$^{-1}$) | Low frequency damping (cm$^{-1}$) | High frequency damping (cm$^{-1}$) | Crossover frequency (cm$^{-1}$) | Frequency Width (cm$^{-1}$) | Carrier concentration 10$^{20}$(cm$^{-3}$) |
|---|---|---|---|---|---|---|---|
| **1.1** | **7.4** | 8063.5 | 4000.0 | 1606.7 | 1127.5 | 770.8 | 2.90 |
| **1.8** | **8.7** | 10308.9 | 1567.6 | 1148.1 | 3756.2 | 164.8 | 4.74 |
| **2.7** | **8.2** | 13308.9 | 4000.0 | 712.0 | 4077.9 | 752.1 | 7.91 |
| **4.8** | **7.4** | 13846.4 | 1915.7 | 869.9 | 5405.3 | 375.6 | 8.56 |
| **7.2** | **7.4** | 15544.9 | 2630.7 | 916.3 | 6439.5 | 451.5 | 10.80 |
| **1.0** | **15.4** | 8822.9 | 1331.7 | 1273.5 | 3394.3 | 4.0 | 3.47 |
| **2.4** | **14.6** | 11374.7 | 1479.2 | 893.9 | 5225.3 | 250.3 | 5.77 |
| **2.8** | **13.3** | 13192.8 | 1339.0 | 1083.9 | 5825.5 | 1000.0 | 7.77 |
| **3.6** | **12.8** | 15204.6 | 1869.8 | 1131.1 | 6262.0 | 344.2 | 10.30 |
| **8.2** | **11.5** | 15804.6 | 2743.0 | 1421.5 | 6935.9 | 325.7 | 11.10 |

**Section S2. Device Assembly and Spectroelectrochemical Measurements**

**Film preparation**. Uniform thin films of Sn:In$_2$O$_3$ NCs were deposited on commercial glass/Sn:In$_2$O$_3$ (50 nm) for *in situ* spectroelectrochemical measurements as well as on Si chips for characterization using a spin-coating method. For spin-coating, substrates were submerged with an adequate volume of NCs dispersion in hexane/octane (1:1 V/V) and spun at 2000 rpm for 1 min, followed by 3000 rpm for 30 s. The NCs assembly were further ligand exchanged with formic acid driven by mass action; films were immersed in 0.1 M formic acid in MeCN for 30 min followed by rinsing multiple times with a mixture of 1:1 MeCN/CHCl$_3$. At the next step, the formic acid ligands were removed by annealing the films at 120 °C for 20 min in a tube furnace under 50 sccm flow of Ar.

**Device assembly.** The NCs films were assembled in a novel *in situ* FTIR spectroelectrochemistry (SEC) device for spectroelectrochemical measurements. The devices were composed of a working electrode (the Sn:In$_2$O$_3$ NC film) and a counter electrode consisting of a glass/Sn:In$_2$O$_3$ substrate coated with Pt nanoclusters. Pt clusters, which served to increase the capacitance of the counter electrode, were deposited by drop-casting Platisol (Solaronix) on the substrates followed by heating in air at 350 °C for 20 min. Electrodes were fused together using a thermoforming Surlyn film (60 μm, Solaronix). A 1.0 cm$^2$ hole in the Surlyn spacer provided a compartment which was further filled with electrolyte (1.0 M tetramethylammonium bis(trifluoromethanesulfonyl)imide (TBATFSI) in tetraglyme). A photograph of the final device is shown in Figure S3.

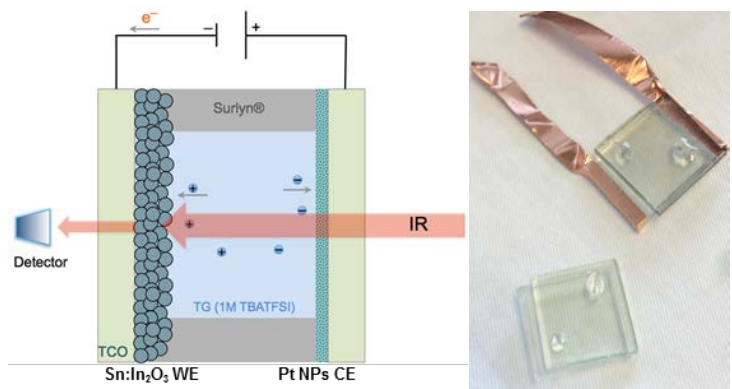

**Figure S3.** Schematic design and a photograph of the in situ FTIR SEC device.

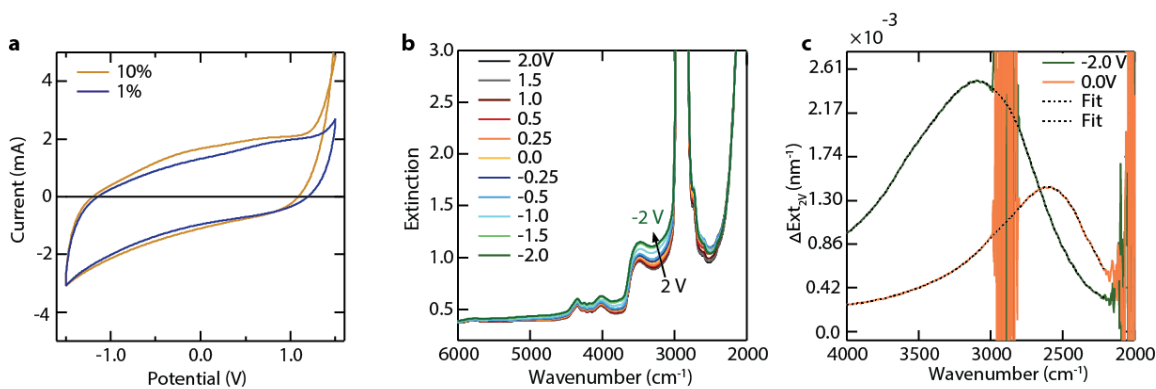

**Figure S4.** a) CV curves of Sn:In$_2$O$_3$ NC films scanned at 20 mV/min. b) A series of in situ FTIR spectra for 1% doped-7.42 nm NCs collected at the indicated potentials. For further analysis, the individual spectra were corrected versus that at 2 V (the oxidized state). Two representative spectra overlapped with their fit curves are shown in (c).

**Electrochemistry.** Electrochemical measurements were made using a S200 Biologic potentiostat. Three cycles of cyclic voltammetry (CV) were run before the *in situ* FTIR measurements. Capacitive CV profiles indicated the high capacitance of the working and counter electrodes (Figure S4a).

The electrochemical charging during *in-situ* FTIR measurements was performed by applying potentials from 2 V to 0.5 V in steps of 0.5 V, 0.5 V to -0.5 V in steps of 0.25 V, and -0.5 V to -1.5 V in steps of 0.5 V. At each potential, optical spectra were taken after 3 minutes of charging time. After three minutes, at all charging potentials, the charging current was less than $10^{-3}$ mA that indicates attainment of an electrochemical equllibirum charging state. Current measured after 3 min is background leakage curent, which had no effect on the optical spectra.

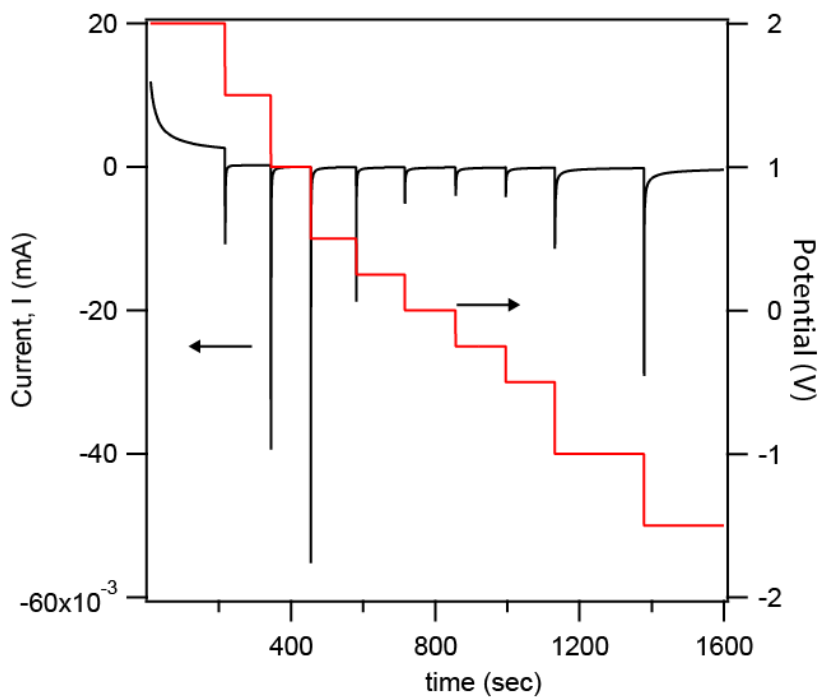

**Figure S5.** Electric potential as a function of time (red curve-right axis) and corresponding charging current as a function of time (black curve-left axis). Electric potential was stepped from 2 V to -1.5 V.

**FTIR Spectroscopy.** Optical transmission spectra were measured using a Bruker Vertex 70 FTIR. For the *in situ* FTIR SEC, the device was attached to a holder with an aperture of ~ 1.1 cm$^2$. The device was titled to a 20 ° angle in order to avoid the internal refection patterns from reaching the detector. A typical *in situ* FTIR spectrum was acquired by backgrounding the detector versus ambient air, then applying a constant potential for 3 min before recording the spectra. A series of sample spectra (raw *in situ* FTIR SEC data) are shown in Figure S4b. The raw FTIR spectra show several different optical features in addition to the LSPR absorption arising from the NC film. Each one of the absorption features results from other components of the electrochemical cell. The most prominent background feature of broad extinction from between 4000-2000 cm$^{-1}$ comes from supporting glass substrates (Figure S6i). In addition to glass, there are broad but weak absorption contributions from the Sn:In$_2$O$_3$ films on both the cathode and anode as well as the very thin Pt nanoparticle film on anode (Figure S6ii-iv). The other prominent features observable in the FTIR spectra in the range of 3500-2500 cm$^{-1}$ come from various vibrational modes of the electrolyte solution (Figure S6v).

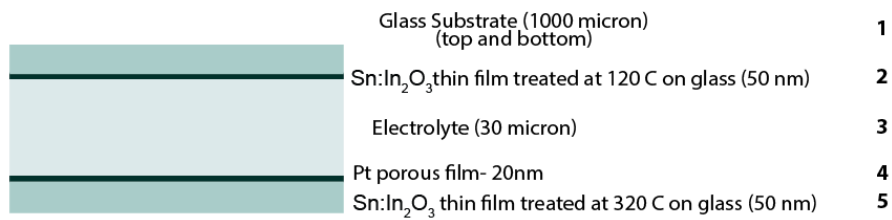
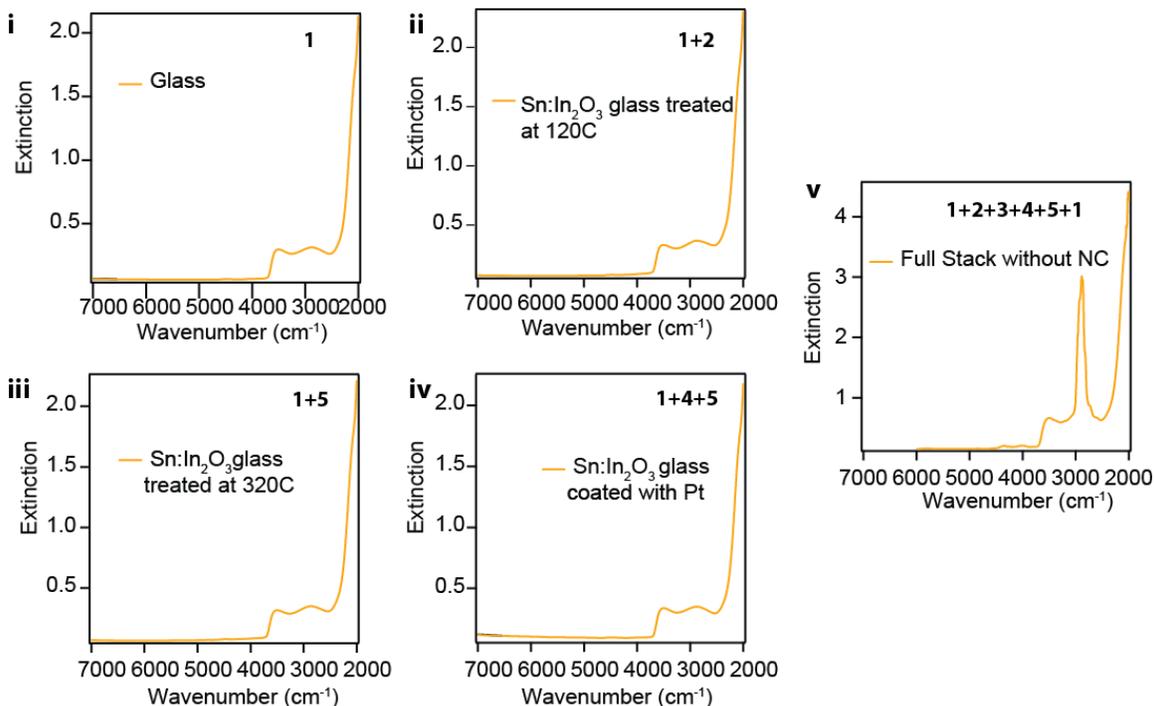

**Figure S6.** Absorption spectra of individual components of the electrochemical cell, supporting glass (i), thin Sn:In$_2$O$_3$ films on cathode and anode after being treated at 120 °C and 320 °C respectively (ii, iii), Pt coated on Sn:In$_2$O$_3$ glass (iv) and full stack without NC film (v).

Solution FTIR spectra for sensitivity analysis were recorded using a thin layer liquid cell attachment. Background spectra were collected using the blank solvents before the NCs dispersion spectral acquisition (more details in Section S6).

**Experimental Data Analysis.** The *in situ* FTIR SEC spectra were further analyzed by correcting for the background and extracting the peak positions and absorption intensity. The spectra were first compared to that recorded at 2 V by subtracting this as a reference spectrum. Using this method, absorption at 2 V is thus zero on our differential plots, and as the potential is stepped negatively a net LSPR absorption appears that is due to injected electrons (Figure S4c).

The question arises as to whether at 2 V the LSPR absorption is negligible and how a nonzero LSPR affects the trend of data shown in Figure 2. Analyzing the raw spectra indicates that the LSPR is completely bleached at 2 V for low doping levels. For high doping levels, however, the LSPR is not completely bleached. This would not cause any misinterpretation due to the following reason. The existence of an LSPR at 2 V is only affecting the qualitative understanding when there is a shift in $\omega_{LSPR}$. For the case of high doing since there is a subtle shift in the $\omega_{LSPR}$ an residual LSPR absorption would only affect the absolute intensity of the corrected spectra, which is not the basis of our interpretations.

To extract the LSPR peak positions, the data were fit using a smoothing function from which the peak positions were extracted. Sample fits are shown in Figure S4c.

**In Situ Electrochemical LSPR Modulation Spectra**. The following three Figures display the full results of LSPR modulation for different doping level and sizes discussed in the main text (Figure 2 and Figure 4a,b).

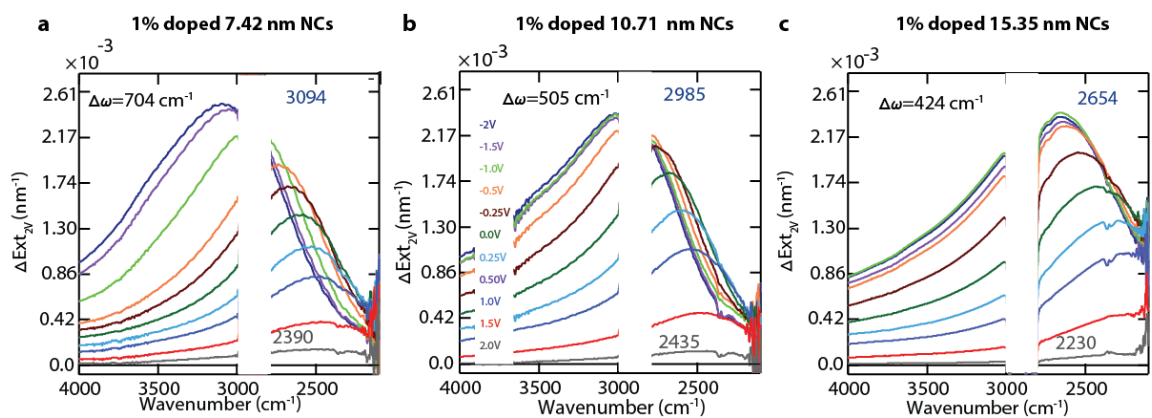

**Figure S7.** Electrochemical LSPR modulation spectra of 1% doped NCs. The peak positions at the most oxidized (1.5 V) and most reduced (-2 V) states are labeled.

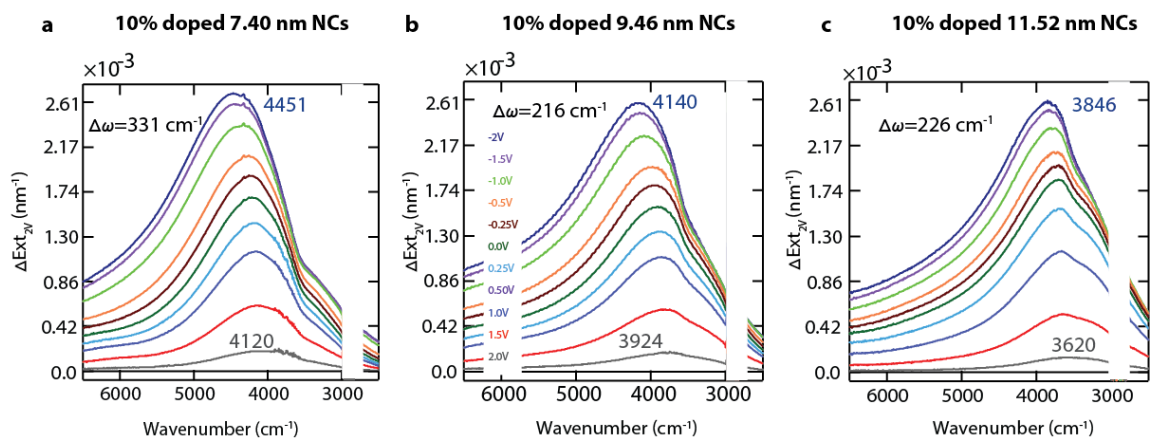

**Figure S8.** Electrochemical LSPR modulation spectra of 10% doped NCs. The peak positions at the most oxidized (1.5 V) and most reduced (-2 V) states are labeled.

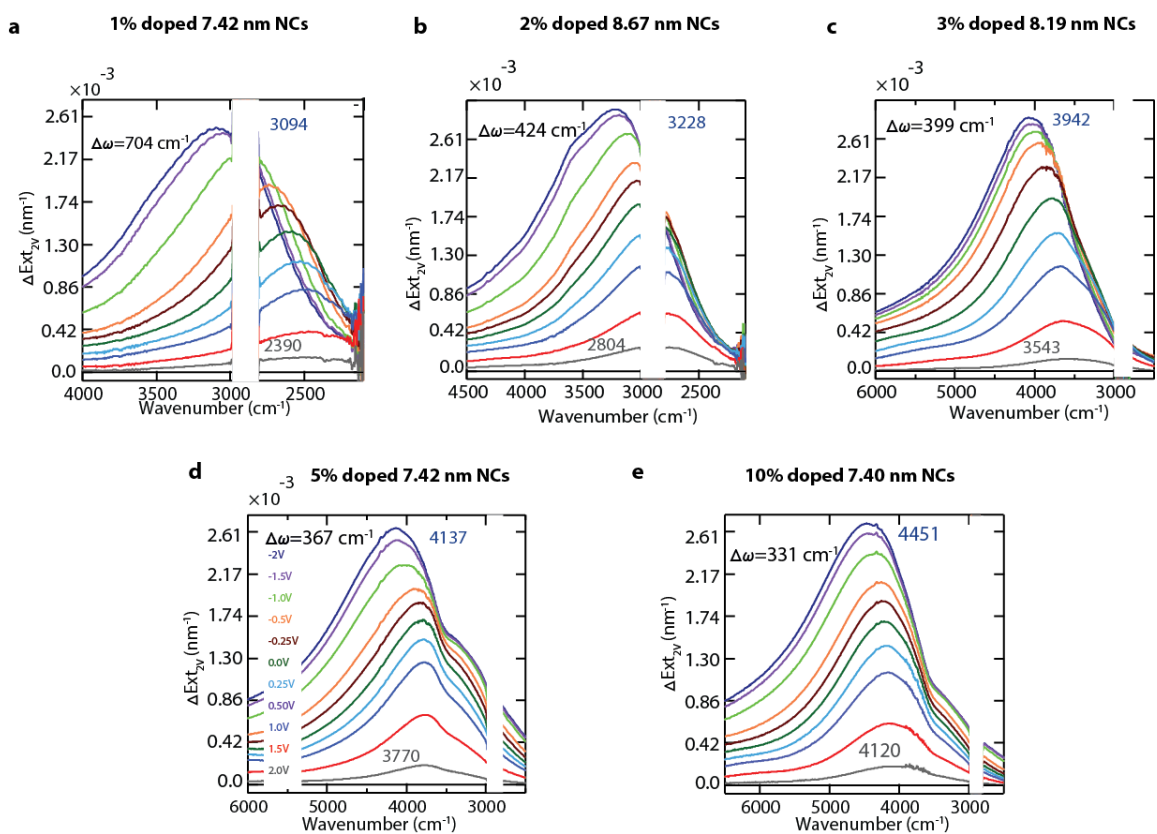

**Figure S9.** Electrochemical LSPR modulation spectra of NCs with various doping levels. The peak positions at the most oxidized (1.5 V) and most reduced (-2 V) states are labeled.

*Modeling Details*

**Section S3. Poisson's Equation**

Poisson's equation was solved numerically for spherical nanocrystals by assuming of uniform dopant concentration and surface potential, $E_S$, using a finite element method. The charge density at any point inside the nanocrystal is made up of mobile electrons, holes and immobile ionized impurity centers.

Here, we have shown the potentials used to solve the Poisson's equation (Figure S8). $E_F$ is the Fermi energy level, $E_{CB}$ is the conduction band minima, $E_{VB}$ is the valence band maxima, $E_I$ is the reference potential and center of the band gap.

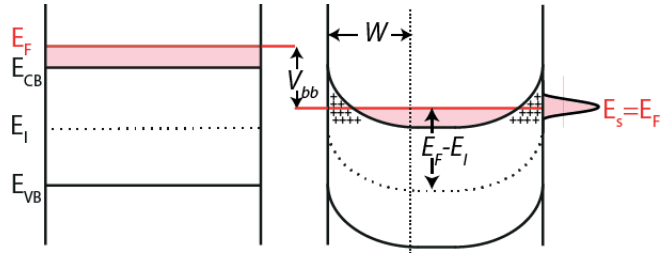

**Figure S10.** The band energetics of a NC under flat band, and band bending conditions. $V_{bb}$ is the band bending potential.

Here, we adapted the dimensionless form of Poisson's equation derived by Seiwatz and Green[2] to solve numerically for a spherical nanoparticle in Cartesian coordinates as,

$$\nabla^2 u = -\frac{e^2 \rho}{\varepsilon \varepsilon_0 kT} \qquad \text{Equation S1}$$

The non-dimensional potential is defined as, $u = \dfrac{E_F - E_I}{kT}$, and k is the Boltzmann constant and T is temperature. $\varepsilon_0$ is the vacuum permittivity, $\varepsilon$ is the static dielectric constant, and $\rho$ is the charge density.

$$\rho = \{\rho_D - \rho_A + p - n\} \qquad \text{Equation S2}$$

where, $\rho_D$ is the donor dopant density, $\rho_A$ is the acceptor dopant density, $p$ is hole density, $n$ is electron density. Here, since we only have aliovalent donor dopants, $\rho_A = 0$

The free electron concentration in the parabolic conduction band is equal to,

$$n = 4\pi \left[ \frac{2m_e kT}{h^2} \right]^{\frac{3}{2}} \left[ F_{\frac{1}{2}}(u - w_{c,I}) \right],$$  Equation S3

where $F_{\frac{1}{2}}(\eta) = \int_0^\infty \frac{x^{\frac{1}{2}} dx}{1 + \exp(x - \eta)}, \; w_{c,I} = \frac{E_{CB} - E_I}{kT}$

Similarly, hole concentration in the parabolic valence band is equal to

$$p = 4\pi \left[ \frac{2m_h kT}{h^2} \right]^{\frac{3}{2}} \left[ F_{\frac{1}{2}}(w_{v,I} - u) \right], \text{ where } w_{v,I} = \frac{E_{VB} - E_I}{kT}$$  Equation S4

If the donor energy level is $E_D$, the activated dopant concentration can be expressed as,

$$\rho_D = \frac{N_D}{1 + 2\exp(u - w_{D,I})} \text{ where } w_{D,I} = \frac{E_D - E_I}{kT}$$  Equation S5

Substituting all the individual terms into Equation S1

$$\nabla^2 u = -\frac{e^2}{\varepsilon \varepsilon_0 kT} \left\{ \frac{N_D}{1 + 2\exp(u - w_{D,I})} + 4\pi \left[ \frac{2m_h kT}{h^2} \right]^{\frac{3}{2}} \left[ F_{\frac{1}{2}}(w_{v,I} - u) \right] - 4\pi \left[ \frac{2m_e kT}{h^2} \right]^{\frac{3}{2}} \left[ F_{\frac{1}{2}}(u - w_{c,I}) \right] \right\}$$

Equation S6

with the boundary condition,

$$u = u_{surf} = \frac{E_{surf} - E_I}{kT}$$  Equation S7

Poisson's equation (Equation S6) was solved in COMSOL using a finite element scheme. Mesh and geometry of the control volume is demonstrated in Figure S9,

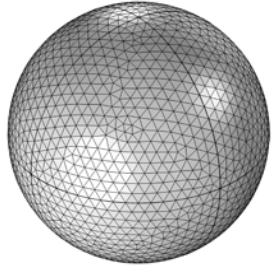

**Figure S11.** Meshed geometry of the spherical nanocrystal of variable radius, R. Poisson's equation was solved over this control volume using a finite element method in COMSOL.

**Section S4. Plasmon Sensitivity to Dielectric change in a core shell model**

LSPR can respond to a change in the dielectric of the surrounding medium, as a result of an oxidized shell or an applied potential. The LSPR spectra of low doped (Figure S10a) and highly doped NCs (Figure S10b) with plasma frequency at 8000 and 15000 cm$^{-1}$, respectively, were simulated in a core shell geometry. It was assumed that for low doped NCs depletion extends 2 nm in from the NC surface at the most oxidizing potential and there is no depletion at the flat band potential. For highly doped NCs, due to their high carrier density, the maximum depletion width was assumed to be 1 nm. In both the cases, a small shift in LSPR frequency was accompanied by substantial intensity modulation. For larger sizes, due to lower modulation in active core volume fraction, the intensity changes were smaller.

Furthermore, to demonstrate the contrasting case where the carriers are depleted uniformly throughout the NC, simulation of extinction spectrum of NCs with different plasma frequencies was performed (Figure S11). Increasing carrier density leads to LSPR shifting towards higher energy, and is accompanied by a substantial increase in extinction intensity. The relationship between LSPR extinction intensity and peak energy is very different if the LSPR is modulated by carrier concentration or surrounding dielectric variations. A given modulation in LSPR intensity would be accompanied by a much greater shift in LSPR energy for a change in carrier concentration, compared to a similar intensity shift from dielectric surrounding (Figure S11). The next section discusses the effects of simultaneous modulations in carrier density and surrounding dielectric on the LSPR.

**The case of low doping level:**

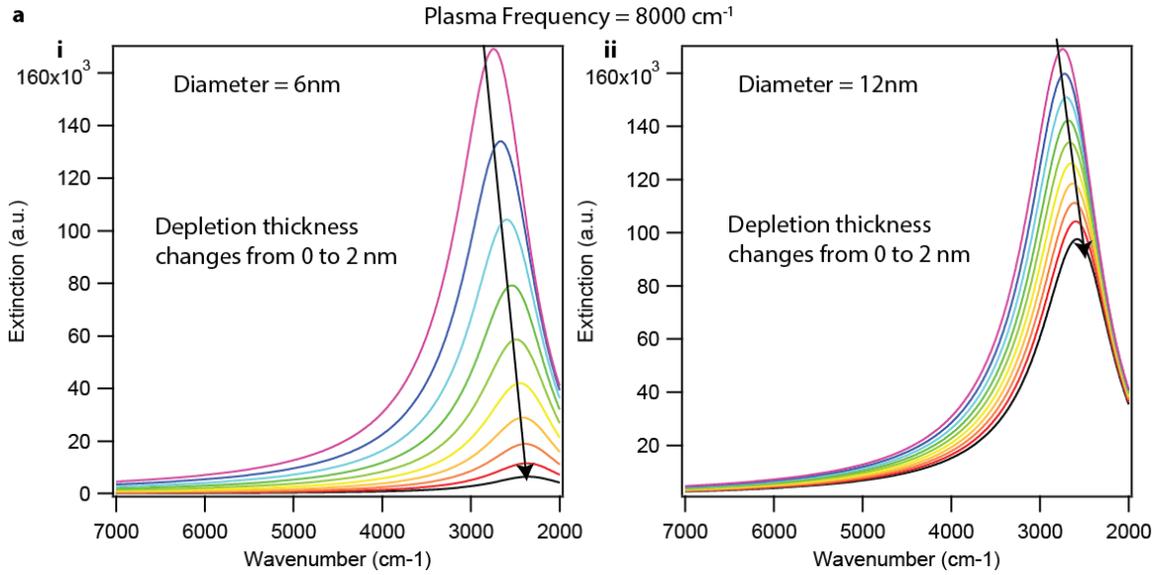

**The case of high doping level:**

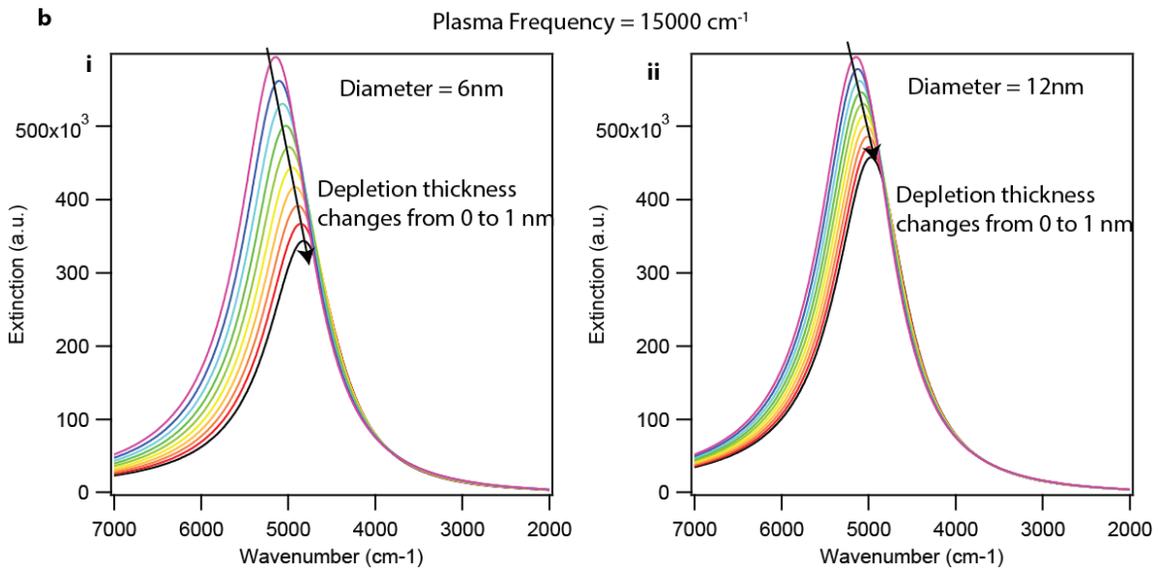

**Figure S12.** a) Dependence of Sn:In$_2$O$_3$ LSPR on the dielectric constant of a depleted shell with variable thickness for the case of low (a) and high (b) doping levels.

**The case of no depletion: pure carrier concentration effect**

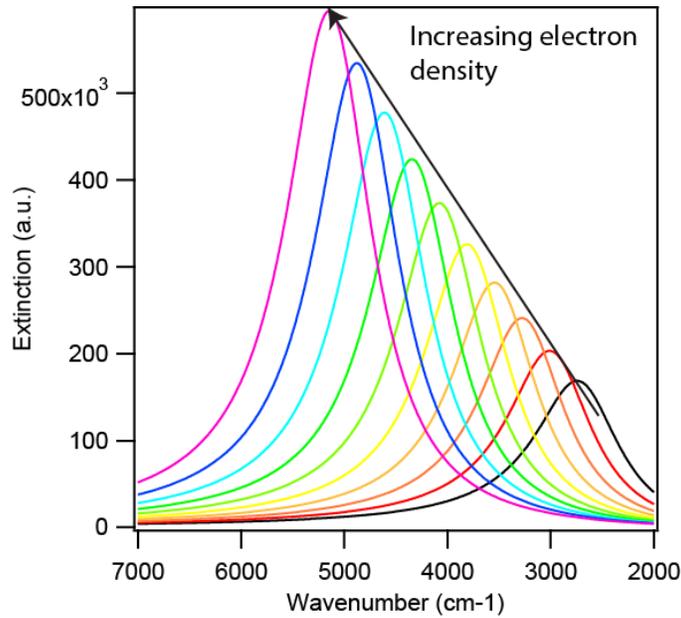

**Figure S13.** Dependence of Sn:In$_2$O$_3$ NC LSPR on the changes in carrier concentration in the Sn:In$_2$O$_3$ NCs with diameter of 10nm assuming no depletion.

**Section S5. Effective Nanocrystal Dielectric Function**

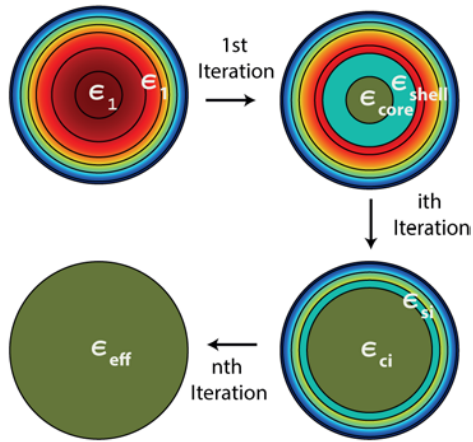

**Figure S14.** Schematic showing the iterative procedure to obtain an effective dielectric function of a depleted NC.

Here, we have developed a method to derive an effective NC dielectric function. This involves applying the Maxwell-Garnet effective medium theory[3] iteratively to a finely discretized sphere into a core and multi-shell configuration (Figure S12). The detailed algorithm to obtain effective dielectric function developed in this work is as follows:

1) Calculated the free electron concentration from the solution of Poisson's equation.
2) Discretized the spherical geometry into a core and 4 shells. Using more than 4 shells did not significantly change our solution. Discretization was based on the carrier density. Six equally spaced focal points in between the maximum and minimum carrier densities obtained from the carrier density profile were calculated
$(n_1 = n_{min}, n_2, ..., n_i, ..., n_5 = n_{max})$.
3) Corresponding to those six carrier concentrations, the radial distance from the center of the sphere was obtained from the carrier profile.
$(r_0 = 0, r_1, ..., r_i, ..., r_5 = R)$
4) To conserve the total number of free electrons within a region (core or shell), volumetrically averaged carrier density was calculated using the following expression $N_i = \dfrac{\int_{r_{i-1}}^{r_i} 4\pi r^2 n_i \, dr}{\int_{r_{i-1}}^{r_i} 4\pi r^2 \, dr}$ where i =1 to 5
5) Using the value of volumetrically averaged carrier concentration obtained from 4, the dielectric function of each region was determined using the Drude model $\varepsilon_i = \varepsilon_\infty - \dfrac{\omega_{p,i}^2}{\omega^2 - \omega\gamma}$ where, $\omega_{p,i} = \sqrt{\dfrac{N_i e^2}{m_e \varepsilon_0}}$ and $\gamma$ is the frequency dependent scattering function expressed as,
$\gamma = \gamma_L - \dfrac{\gamma_L - \gamma_H}{\pi}\left(\tan^{-1}\left(\dfrac{\omega - \gamma_X}{\gamma_W}\right) + \dfrac{\pi}{2}\right)$ where, $\gamma_L$ is the low frequency damping, $\gamma_H$ is high frequency damping, $\gamma_X$ is a crossover frequency, and $\gamma_W$ is the crossover width. All the parameters other than carrier concentration are taken as uniform throughout the nanocrystal.
6) Now starting from the core and first inner shell the discretized nanocrystal has the dielectric function $\varepsilon_{core} = \varepsilon_1$ and $\varepsilon_{shell} = \varepsilon_2$ (Figure S12). Using core-shell Maxwell Garnett effective medium theory, an effective dielectric function ($\varepsilon_{core,i}$) of the new core is calculated. Given a core radius of a= $r_1$ and shell outer radius of b=$r_2$,

$$\varepsilon_{core,new} = \varepsilon_{shell} \dfrac{a^3(\varepsilon_{core} + 2\varepsilon_{shell}) + 2b^3(\varepsilon_{core} - \varepsilon_{shell})}{a^3(\varepsilon_{core} + 2\varepsilon_{shell}) - b^3(\varepsilon_{core} - \varepsilon_{shell})}$$

Equation S8

7) The above equation was applied again with the new $\varepsilon_{core,new}$ and $\varepsilon_{shell,new}$, *i.e.*, the next inner most shell to the new core (Figure S12). This procedure was repeated until we obtained a single effective dielectric function for the NC, $\varepsilon_{p\_effective}$ (Figure S12).
8) Using the effective dielectric function, absorption of the single NC was calculated *via* Mie theory.

**Section S6: Effective film dielectric function and device optical modeling**

The sandwich cell designed in-house for *in situ* spectroelectrochemistry was modeled using commercial SCOUT software (www.wtheiss.com) by employing several built-in optical models, including the T-matrix method to calculate extinction spectrum and the Drude model to express the dielectric function of $Sn:In_2O_3$ thin films on a glass substrate. To model the device the following steps were followed:

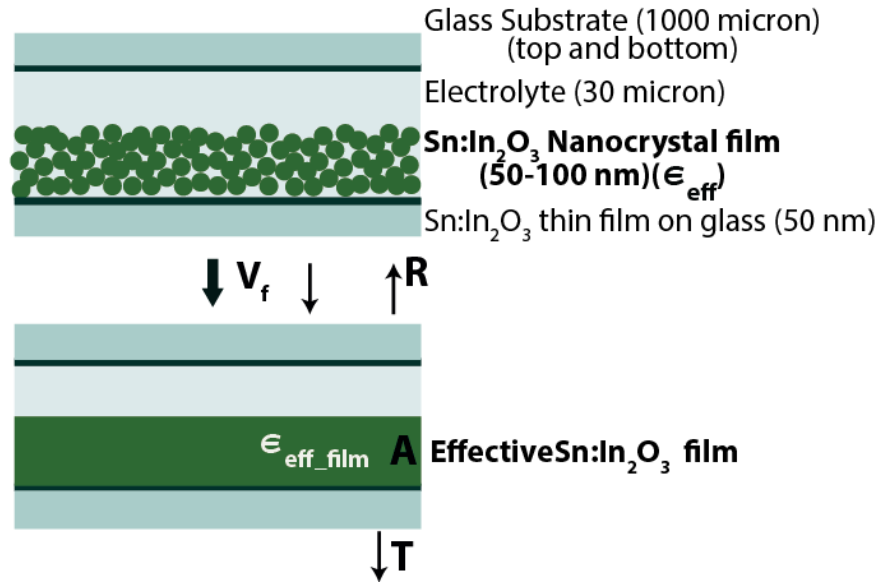

**Figure S15:** Schematic of the modeled sandwich device configuration. Here, an $Sn:In_2O_3$ NC film with effective dielectric function $\varepsilon_{eff}$ is coated on the top of an $Sn:In_2O_3$ thin film coated glass substrate. To model the $Sn:In_2O_3$ NC film, the effective dielectric function of the film was calculated using the Maxwell-Garnett equation.

1) Building up the material library, which consists of glass substrate, $Sn:In_2O_3$ thin film, $Sn:In_2O_3$ nanocrystals and electrolyte.

2) Defining the dielectric model for each of material.

3) The glass substrate was modeled as the sum of several Lorentians. The parameter for each Lorentian was obtained by fitting the experimental extinction spectrum with a modeled one. Similarly, the spectrum of an $Sn:In_2O_3$ thin film on glass was fitted to obtained the Drude parameters for the background substrate used in our measurements.

4) Effective dielectric function of the NCs was calculated using the procedure described in the previous section and imported into SCOUT.

5) Effective dielectric function of the NC film consisting of NCs with a fixed volume fraction and electrolyte filling the pores was calculated using the built in Maxwell Garnett effective medium model.

6) Sample extinction spectra for a given doping level and size of the NCs for different applied potential obtained from this procedure (Figure S13) are analogous to those of experimentally measured spectra (Figure S4). To obtain the difference spectra, similar to our experimental data analysis procedure, all spectra were background subtracted against the spectrum at the most oxidized potential (Figure S14).

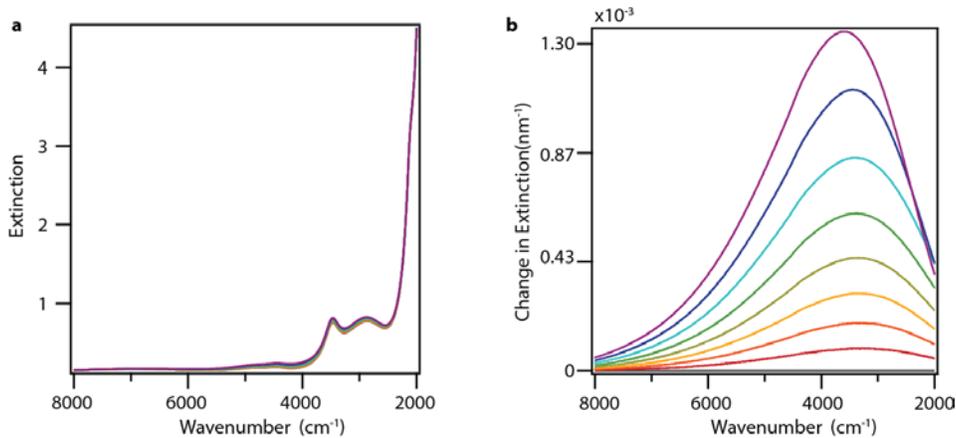

**Figure S16.** A series of modeled spectra for 2% doped-6nm NCs evaluated using the NC dielectric functions calculated at several different surface potentials (a). For further analysis, the individual spectra were subtracted versus that of the most oxidized state (b).

**Section S7. Validity of the optical models**

The optical models adapted in this work are based on solution of continuum Maxwell equations. The accuracy and limitations the optical models are discussed in this section.

1) Optical properties of the individual NCs

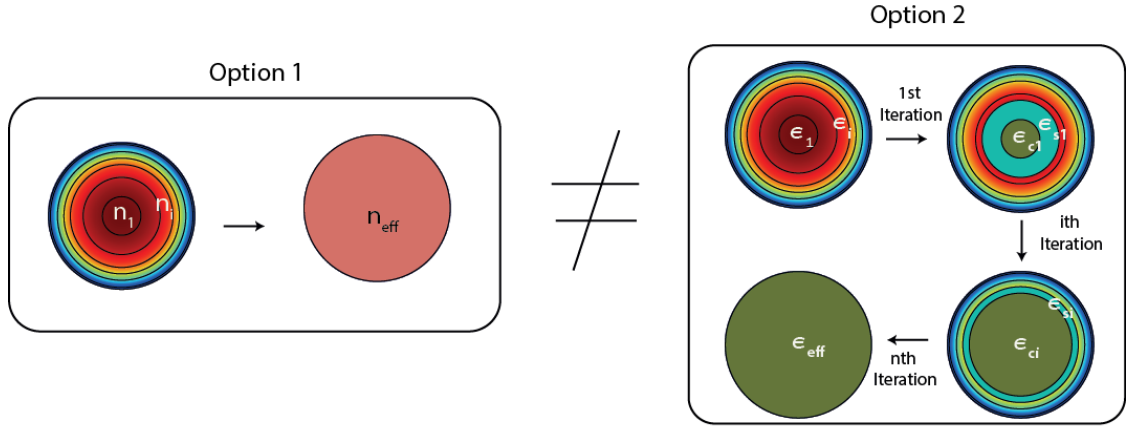

**Figure S17.** Schematics depicting two possible volume averaging methodologies to obtain the effective dielectric permittivity of the NC, first, by calculating the volume averaged carrier concentration (left) and second by discretizing the NC into given number of shells and a core, followed by effective permittivity calculation using an iterative core-shell approximation. The two methodologies are not equivalent and indicate very different optical extinction properties.

Optical properties of individual NCs were calculated using a core-shell model. As explained in Section S5, based on the radial carrier distribution obtained from the solution of Poisson's equation, a NC is discretized into a system consisting of a core and five shells. The permittivity of each of these domains was calculated using volume-averaged carrier density within each of the domain. The effective permittivity calculated using an analytical core-shell model that takes into account the interaction between each of the discretized domains within the NC. This core-shell approach to find effective permittivity is distinct from choosing an effective (uniform) carrier concentration, which is another common treatment in literature[4–7]. We emphasize that the effective permittivity obtained from the core-shell model is not a Drude type permittivity and therefore cannot be expressed using an effective carrier concentration and damping constant. Moreover, to test the accuracy of the core-shell approximation we simulated the extinction spectra based on three different treatments, namely using a Drude type permittivity from an effective carrier concentration, effective permittivity from the core-shell approximation, and numerical solution of Maxwell equations using radially varying carrier concentration. The extinction spectrum calculated using an effective carrier concentration was at lower energy and had a distinct shape compared to one obtained through full numerical simulation. On the other hand, the extinction spectrum calculated through the core-shell approximation was identical to the extinction spectrum obtained through full simulation. The agreement between the simulation and core-shell approximation reaffirms the validity of the core-shell model used in our study.

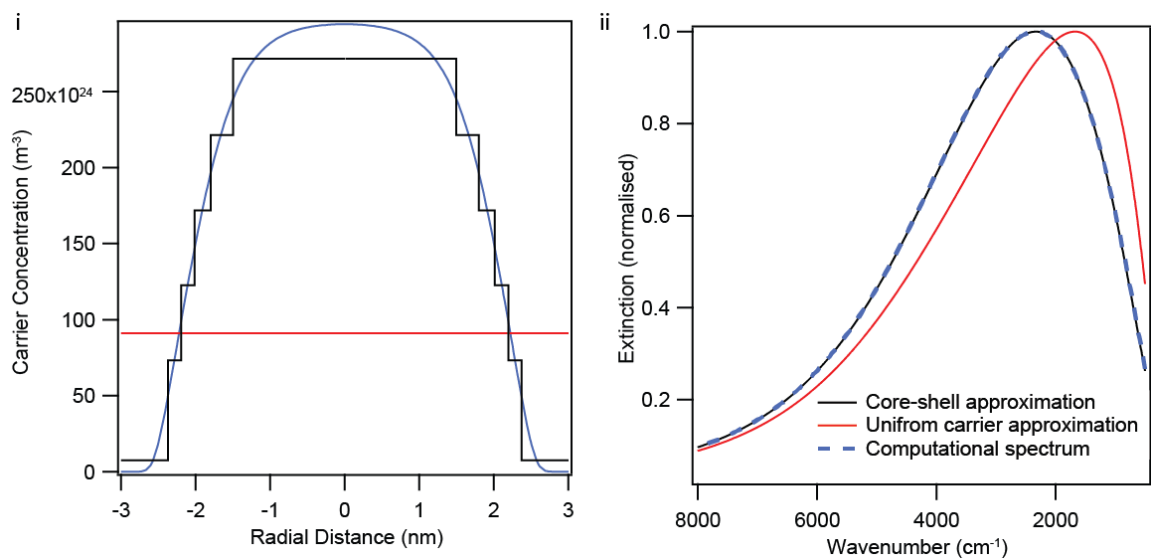

**Figure S18** i) The three different carrier concentration profiles used to calculate the extinction properties of a NC: first, a radial carrier concentration distribution obtained by solving Poisson's equation for a low doped NC (blue); second, the continuous radial distribution was discretized into a core and five shells (black); and third, volume averaged uniform carrier concentration (red). ii) Extinction spectra corresponding to each of the carrier profiles in (i). Extinction based on the uniform distribution was calculated using Mie theory (red), extinction based on the discretized carrier distribution using the core-shell approximation (black) and extinction of the NC with the continuous radial distribution using full field electromagnetic simulation (blue dashes). An excellent match was obtained between the core-shell approximation and full field simulation.

2) Error originating from possible quantum effects.

The optical analysis in our study is based on the classical theory of electromagnetism. Furthermore, the calculated band profiles were based on Poisson's equation that again does not take into account any quantum effects. There are two major cases where the quantum effects can become relevant as applied to NC LSPR,

a) Quantum size effects: There are two possible origins of quantum size effects in our system, namely NC size and depletion region thickness. The smallest average NC size used in our study was 7.42 nm. At this size, quantum size effects, if any, are minimal[8,9].

The depletion region, due to its very low carrier density, has optoelectronic properties like a dielectric. Even though in some cases, the depletion region is less than 1 nm thick, where the quantum or surface effects may be anticipated, the optical properties of dielectric materials have been shown to be substantially unchanged in this quantum size regime. Moreover, the presence of native oxide coatings of around a nanometer thick is well documented in metal nanoparticles such as Al and Ag and conventionally such oxide layers are given classical dielectric treatment while calculating the optical properties of such metal nanoparticles[10–12].

b) Low carrier density: At ultra-low carrier density ($<10^{19}$ cm$^{-3}$) the origin of infrared absorption transitions from collective excitation or LSPR to quantum mechanical transitions. This cross-over is primarily observed in NCs with less than 2-3 electrons per NC. LSPR in semiconductor NCs has been demonstrated with handful of electrons, e.g. in ZnO NCs with as little as 4 e- per NC[13,14]. The minimum doping level among all the NCs used in this study is 1.1% and the corresponding carrier concentration estimated via Drude fitting is 2.90x10$^{20}$ cm$^{-3}$. This means a 7.4 nm NC contains approximately 62 electron, which is at least an order of magnitude more than the quantum limit.

3) Validity of Maxwell-Garnet approximation

Maxwell-Garnet is a simplistic mean field model that does not take into account strong near field interaction between particles. The Maxwell-Garnet approximation may be expected to fail if the NCs are touching or very close to each other. The cutoff separation (s) between NCs below which the approximation strongly diverges from experimental observations depends on the ratio of dielectric permittivity of the particle ($\epsilon_p$) to that of surrounding medium ($\epsilon_m$) as well as size of the particle (a). In a theoretical study, Reich and Shklovskii[15] explored the divergence of effective permittivity from the approximate mean field models including Bruggeman and Maxwell-Garnett for materials with different permittivity and particle size. The authors suggested that if the ratio, $\epsilon_p/\epsilon_m$, is less than 30 (for materials like indium oxide), than whenever ratio between the separation and particle size (s/a) is greater than 0.025, Maxwell-Garnett predicts effective permittivity with high accuracy than Bruggeman. The authors also showed that the accuracy of Maxwell-Garnet substantially decreases with increasing $\epsilon_p/\epsilon_m$ (for materials like PbSe, and PbS).

In our study, the dielectric constant of the particle is less than 10, and largest particle size is 14 nm (a=7 nm). As shown by our simulation (Figure S22), the minimum depletion width among all the cases in this study is for highest doping level under reducing conditions. The depletion width in this condition is still greater than 0.25 nm. These data suggest that s/a ratio for all cases in our study is always greater than 0.036. Other particle sizes, potentials, and doping levels yield larger values of s/a, suggesting that Maxwell-Garnet should always be able to predict the effective optical properties of our NC films with high accuracy. The conclusion of the study by Reich and Shklovskii[15] applied to our system gives us high degree of confidence in our simulation results.

**Section S8. Modeled electrochemical LSPR modulation spectra**.

The multi-scale modeling approach described in the last 3 sections was used to model the extinction spectra as a function of applied potential for different doping levels and sizes of NCs. Below, we summarize the results of our modeling work. The first column of Figures S19-S21 shows band profiles obtained by solving Poisson's equation. Solid lines shows the conduction band minima and dashed line of same color shows the corresponding Fermi level. The second column shows the corresponding carrier concentration profiles. The third column shows the corresponding absorption of a single NC calculated using the effective dielectric function of the NC and the fourth column shows the extinction difference spectra of the sandwich device.

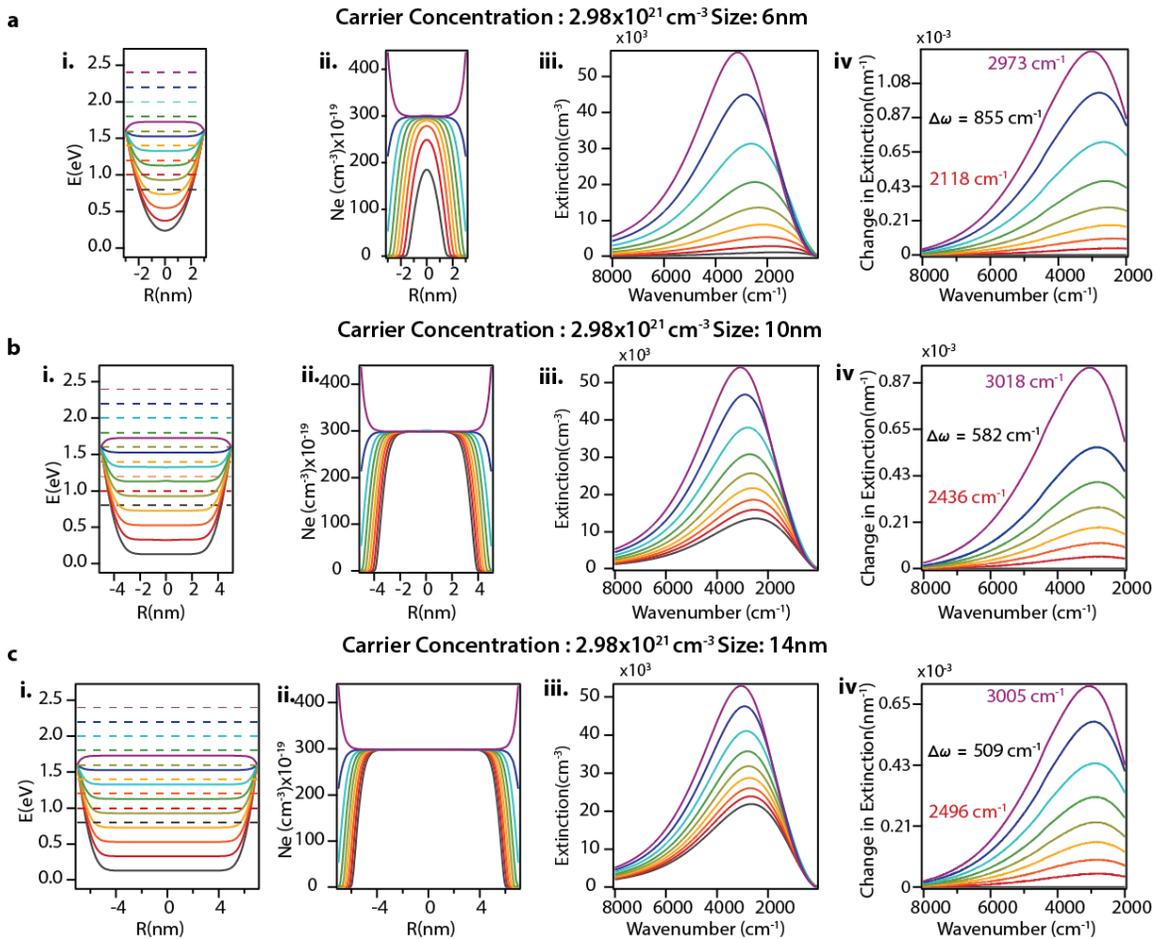

**Figure S19.** Size effect on LSPR modulation for low-doped NCs was modeled in three steps. First, Poisson's equation was solved for band bending profiles (i) and carrier

concentration profiles (ii). Radial carrier concentration profiles corresponding to different surface potentials were used to model absorption spectra of single NCs (iii). Taking into account the film and device effects, optical modeling of single NCs was extend to model the optical device (iv). Frequency shift was maximum for smaller NCs, which is explained by the carrier concentration profiles wherein there is substantial change in carrier concentration in the core for small NCs, while changes in concentration are largely limited to the shell for larger NCs.

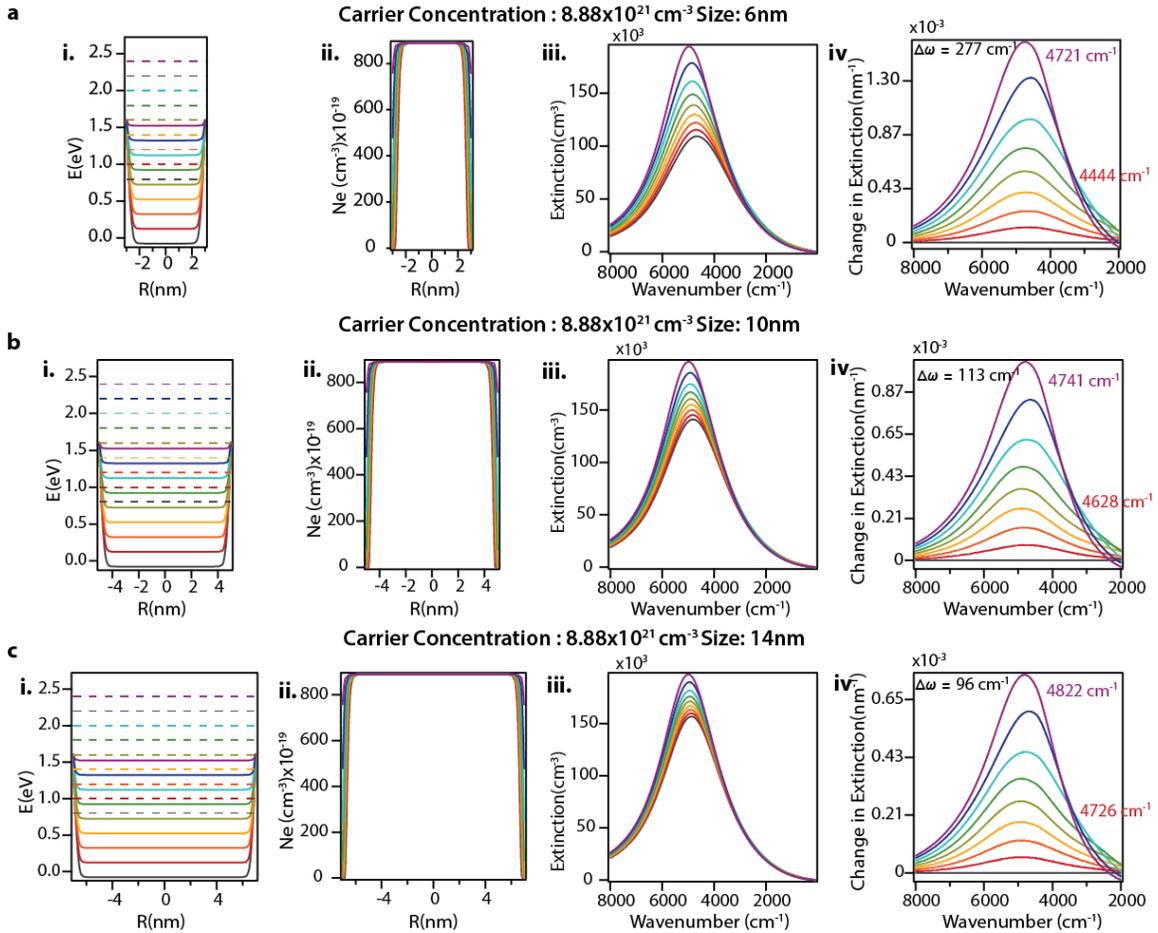

**Figure S20.** Size effect on LSPR modulation for highly doped NCs was modeled in three steps. The procedure was as described for low doped NCs. Frequency shift is relatively small for highly doped NCs since, as is evident in carrier concentration profiles, the charging is largely limited to within 1 nm of the surface of the NCs.

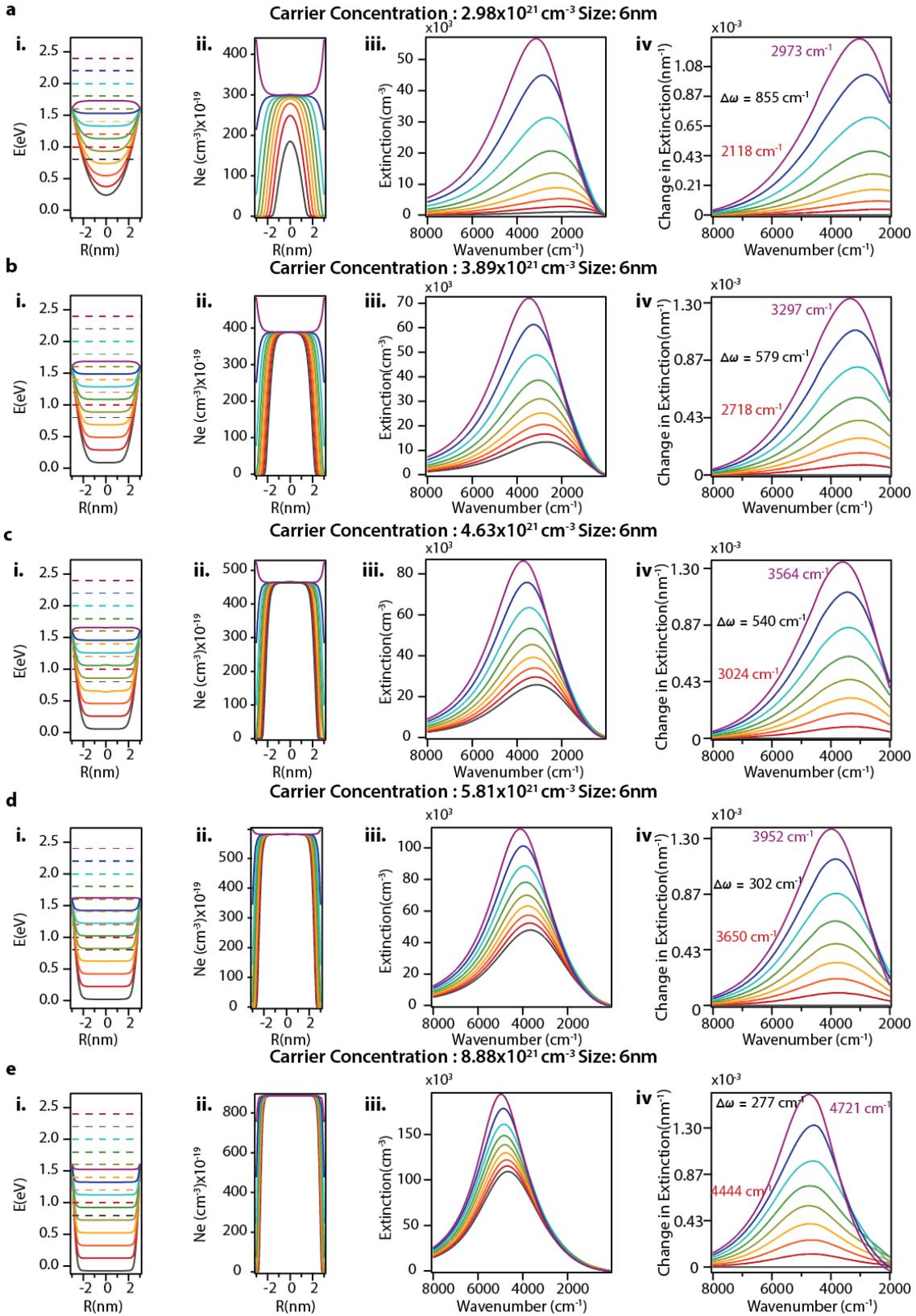

**Figure S21.** Doping concentration effect on LSPR modulation for 6 nm NCs. The procedure as described for low and highly doped NCs was followed here as well. LSPR shift decreases with increasing doping concentration. From the concentration profile as a function of applied surface potential, it can be seen that for low doping concentration there is a substantial change in carrier concentration throughout the NC, but as the doping level increases, charging is mainly limited to the NC surface layer.

**Section S9. NC Plasmon Sensitivity Analysis**

**Experiment.** Sensitivity of LSPR of Sn:In$_2$O$_3$ NCs to the changes in refractive index of the surrounding medium was measured by collecting extinction spectra of NCs dispersed in several solvents such as hexane, toluene, chloroform, and tetrachloroethylene. FTIR spectra were collected using a Bruker Vertex 70 FTIR. Solution ensemble measurements were performed using a FTIR liquid cell with a 0.25 mm path length. LSPR peak position was extracted by fitting the spectra to the sum of several Lorentian functions. As per LSPR condition from Mie theory, and substituting dielectric value using Drude function, we obtain, $\lambda_{LSPR}^2 = Sn^2$ where, $\lambda_{LSPR}$ is the peak LSPR wavelength, $n$ is the refractive index of the surrounding, and $S$ is the sensitivity factor. This relation suggests that the slope of the curve $\lambda_{LSPR}^2$ vs. $n^2$ would give us the sensitivity factor. Following this equation, we plotted $\lambda_{LSPR}^2$ vs. $n^2$ for NC with different doping level and size and obtained the sensitivity factor for each one of them.

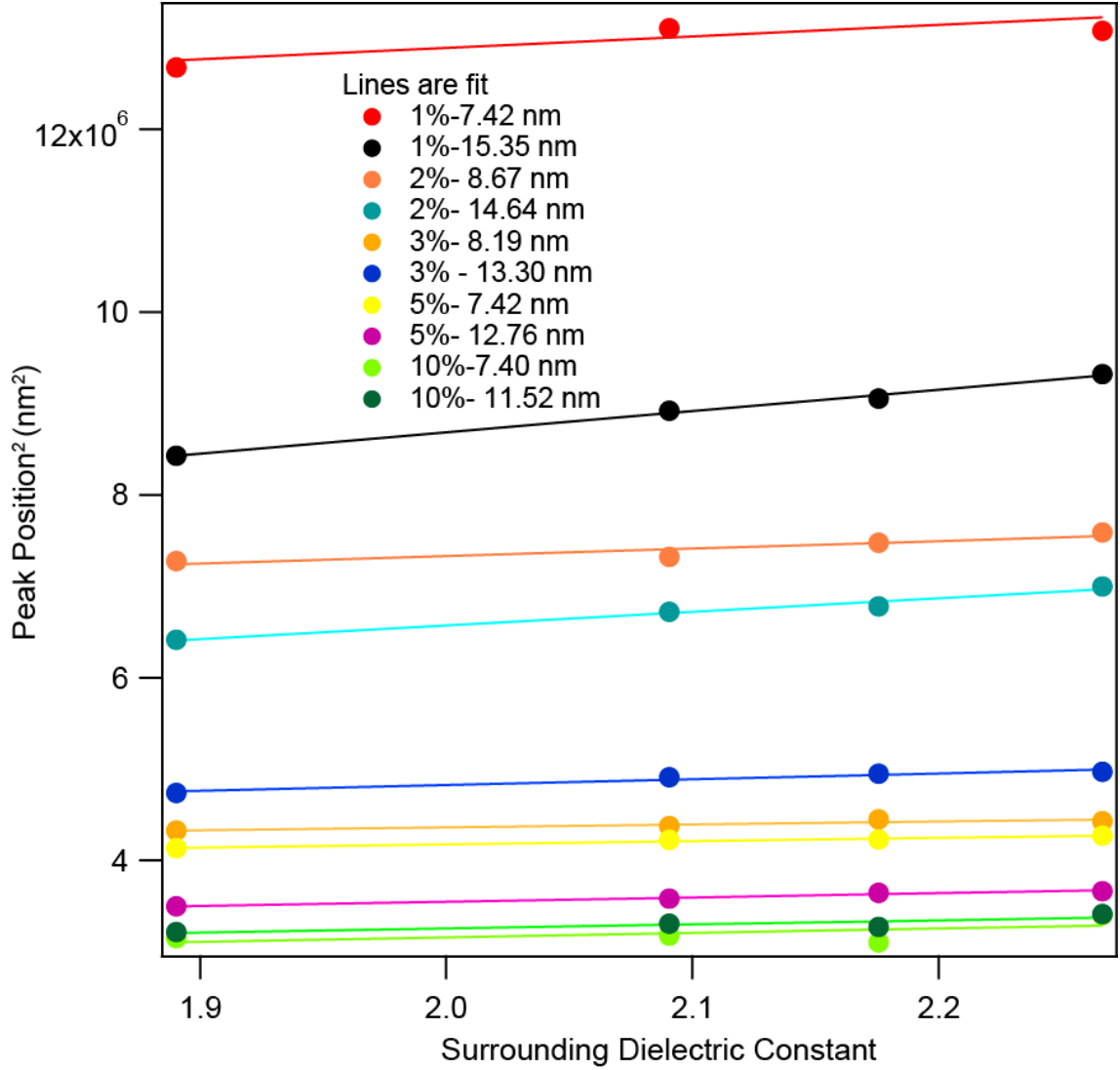

**Figure S22.** $\lambda^2_{LSPR}$ vs. $n^2$ curves plotted for different doping levels and sizes. The linear slope of each curve was found to evaluate the sensitivity of LSPR to the surroundings.

**Model.** In order to model the theoretical sensitivity of the depleted NCs, we used the effective dielectric function of a NC to predict the absorption of the NC using Mie theory (Equation S9)

$$C_{abs}(\omega) = 4\pi R^2 (k(\epsilon_H)^{1/2} R) \text{Im} \left\{ \frac{\epsilon_p(\omega) - \epsilon_H}{\epsilon_p(\omega) + 2\epsilon_H} \right\}$$

Equation S9

Absorption profiles were simulated for refractive indices varying from 1.3 and 1.6, and using these profiles, LSPR peak wavelength was found as function of refractive index for various NCs. To extract the sensitivity of the NC, same procedure was followed as in the experimental analysis, using instead the theoretical LSPR peak position.

**Section S10. Near-field Enhancement.**

Using the concentration profile obtained from Poisson's equation (here, we chose 1.4 eV as the surface potential), the spatially varying dielectric function was derived using the Drude model. A spherical NC was surrounded by a sphere representing the surrounding medium (*i.e.*, air (*n*=1)). This whole system was then surrounded by a perfect index matching layer which prevented unwanted reflections from the outside boundary. The maximum and minimum mesh sizes in the NC were set to 4 and 0.01 nm, respectively. This ensures fine meshing, yielding typically 2 million degrees of freedom, which corresponds to 3 – 15 GB of RAM when using the direct PARDISO or MUMPS solver. Maxwell's equations were solved in the scattered field formulation with the background electric field propagating along Z-axis and polarized along X-axis.